\documentclass{article}

     \usepackage[preprint, nonatbib]{neurips_2025}

\usepackage[utf8]{inputenc} 
\usepackage[T1]{fontenc}    
\usepackage{hyperref}       
\usepackage{url}            
\usepackage{booktabs}       
\usepackage{amsfonts}       
\usepackage{nicefrac}       
\usepackage{microtype}      
\usepackage{xcolor}         
\usepackage{mathtools}
\usepackage{physics}
\usepackage{multirow}
\usepackage{tabularx}
\usepackage{adjustbox}
\usepackage{amsmath}

\usepackage[backend=bibtex]{biblatex}
\usepackage[normalem]{ulem}
\usepackage{caption}
\usepackage{subcaption}
\usepackage[title]{appendix}
\title{Edge-ASR: Towards Low-Bit Quantization of Automatic Speech Recognition Models}

\let\SUP\textsuperscript
\author{%
    Chen Feng\SUP{1},
    Yicheng Lin\SUP{2}, 
    Shaojie Zhuo\SUP{3},
    Chenzheng Su\SUP{4}, \\
    Ramchalam Kinattinkara Ramakrishnan \SUP{5},
    Zhaocong Yuan \SUP{6}, 
    Xiaopeng Zhang\SUP{7}\\
    \\
    Qualcomm AI Research \thanks{Qualcomm AI Research is an initiative of Qualcomm Technologies, Inc.} \\
    \texttt{\{\SUP{1}chenf, \SUP{2}yichengl, \SUP{3}shaojiez, \SUP{4}chenzhen, \SUP{5}rkinatti, \SUP{6}zhaocong, \SUP{7}xiaopeng\}}\\
    \texttt{@qti.qualcomm.com}
}

\begin{document}

\maketitle

\begin{abstract}

Recent advances in Automatic Speech Recognition (ASR) have demonstrated remarkable accuracy and robustness in diverse audio applications, such as live transcription and voice command processing. However, deploying these models on resource-constrained edge devices (e.g., IoT device, wearables) still presents substantial challenges due to strict limits on memory, compute and power. Quantization, particularly Post-Training Quantization (PTQ), offers an effective way to reduce model size and inference cost without retraining. Despite its importance, the performance implications of various advanced quantization methods and bit-width configurations on ASR models remain unclear. In this work, we present a comprehensive benchmark of eight state-of-the-art (SOTA) PTQ methods applied to two leading edge-ASR model families, Whisper and Moonshine. We systematically evaluate model performances (i.e., accuracy, memory I/O and bit operations) across seven diverse datasets from the open ASR leader-board, analyzing the impact of quantization and various configurations on both weights and activations. Built on an extension of the LLM compression toolkit, our framework integrates edge-ASR models, diverse advanced quantization algorithms, a unified calibration and evaluation data pipeline, with detailed analysis tools. Our results characterize the trade-offs between efficiency and accuracy, demonstrating that even $3$-bit quantization can succeed on high capacity models when using advanced PTQ techniques. These findings provide valuable insights for optimizing ASR models on low-power, always-on edge devices.


\end{abstract}

\section{Introduction}
\label{sect:intro}

Real-time Automatic Speech Recognition (ASR) models have revolutionized the natural audio processing pipeline and achieved remarkable accuracy and robustness in diverse applications such as live transcription, voice command recognition and assistive use cases (\cite{ASRSurvey}). As interest in on-device ASR continues to grow, there is an emerging demand for models that operate efficiently on low-power, always-on edge hardware (e.g., IoT device, wearable). Among the state-of-the-art (SOTA) edge-ASR models, OpenAI's Whisper (\cite{Whisper}) has demonstrated strong accuracy in compact model sizes, making it a practical candidate for edge inference. Meanwhile, Useful Sensors' Moonshine (\cite{Moonshine}) reduces latency and adapts well to variable-length audio inputs, which is further optimized for resource-constrained deployment. Both model families currently lead the open ASR leader-board (\cite{Leaderboard}) across seven evaluation datasets, among model sizes ranging from $27$M to $244$M parameters. 

Despite the advances, deploying ASR models on edge devices remains challenging due to memory, compute and power constraints. Most existing low-power neural processing units (NPUs), micro controllers and digital signal processors (DSPs) are designed as fixed-point inference accelerators, making full precision models impractical. Quantization, particularly Post-Training Quantization (PTQ), addresses this by mapping weights and activations to low-bit data formats without retraining, directly reducing model size, latency and power consumption while maintaining acceptable accuracy. Its simplicity and hardware compatibility have driven widespread industry adoption. 

While many advanced PTQ techniques have been proposed and studied in the context of large language models (LLMs) (\cite{QuantLLM}), their applications to edge-ASR remain unexplored. Existing quantization toolkits and benchmarks (\cite{LLMC}) focus on LLM model architectures, and task-specific datasets that do not apply to speech recognition directly. Furthermore, although the open ASR leader-board provides standardized full precision baselines, it lacks coverage of quantized models, leaving the performance trade-offs from quantization in this domain largely unexplored. 

\begin{figure}[!t]
  \centering
  \includegraphics[width=1.0\linewidth]{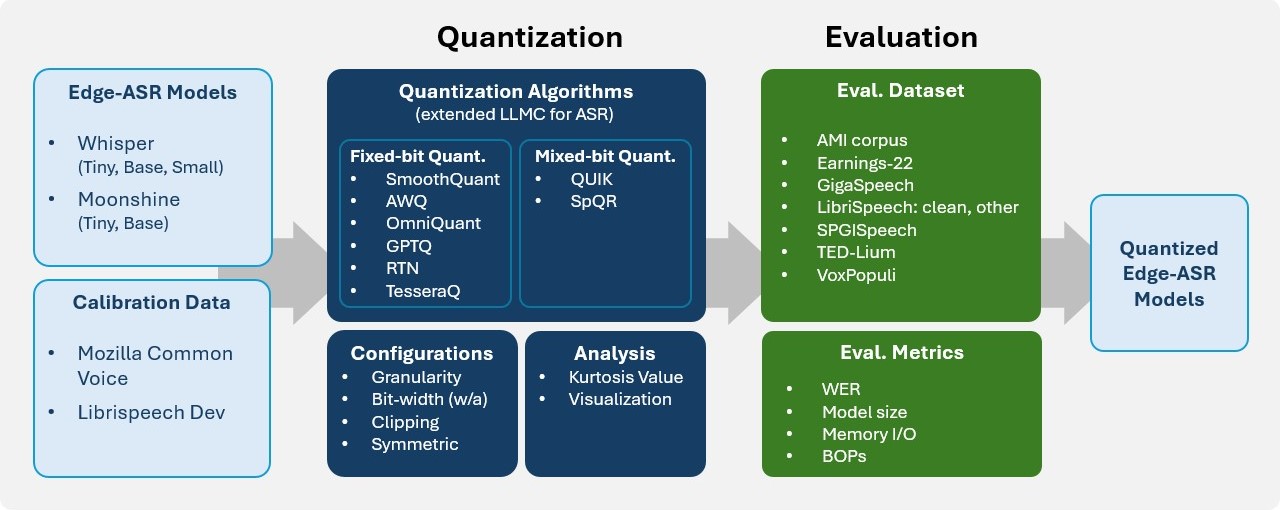} 
 \caption{An overview of our edge-ASR benchmark workflow, which incorporates edge-ASR models, diverse quantization algorithms extended from LLM compression toolkit, unified calibration and evaluation data pipeline, and analysis tools.}
  \label{fig:overview}
\end{figure}

In this work, we fill this gap by benchmarking eight PTQ methods applied to two Transformer-based edge-ASR families, Whisper and Moonshine. The quantization methods are categorized into three primary strategies: transformation-based, reconstruction-based and rounding optimization. We systematically evaluate quantization of weights and activations across seven public ASR datasets, measuring accuracy, memory I/O and bit operations (BOPs) (\cite{BOP}). We also perform ablation studies on calibration data sources and sizes, and analyze quantization sensitivity. All experiments are built on an extension of the LLM compression toolkit (\cite{LLMC}), enabling a reproducible and extensible ASR quantization framework. The overall workflow is demonstrated in Figure \ref{fig:overview}.

\textbf{Contributions.} \textbf{(a)} We deliver the first systematic benchmark of quantized edge-ASR models, assessing eight advanced PTQ methods over multiple datasets, architectures and configurations, to establish a quantized standard for the open ASR leader-board. \textbf{(b)} We extend and contribute to the LLM compression toolkit to support ASR models, enabling an end-to-end, extensible quantization workflow for speech recognition. \textbf{(c)} Through extensive evaluations, analysis and comprehensive comparisons from various factors (e.g., model accuracy, memory I/O, BOPs), we provide valuable insights into how different quantization methods and configurations impact model performance, offering guidance for deploying ASR on memory and compute constrained low-power edge devices. 

\section{Preliminaries}
\label{sect:preliminaries}

\subsection{Quantization Algorithms}
Quantization maps high-precision floating-point values to discrete low-precision levels, reducing memory footprint and accelerating inference. We focus on integer-uniform Post-Training Quantization, which is broadly supported by edge hardware accelerators. The general quantization formula for weights or activations is given by:
\begin{equation}\label{eq:quant}
\begin{aligned}
x_{q} &= clip \left(\lfloor\frac{x_{f}}{\Delta}\rceil + z, N_{min}, N_{max}\right)  \\
\end{aligned}
\end{equation}
where $x_f$ denotes the floating-point value, $x_q$ denotes the low-precision quantized value. $\Delta$ and $z$ denote the scaling factor and zero-point, respectively. $[N_{min}, N_{max}]$ represents for the quantized range based on the target bit-width $n$. For symmetric quantization, the zero-point $z$ is set to zero. 

Post-Training Quantization algorithms can be broadly categorized into three primary strategies: scaling-based transformation, reconstruction-based optimization, and rounding-based optimization. \textit{Scaling-based transformation} techniques, such as SmoothQuant (\cite{SmoothQuant}), AWQ (\cite{AWQ}), and OmniQuant (\cite{OmniQuant}), apply learned or heuristic scaling to weights and activations, reshaping their distributions to be more quantization-friendly. \textit{Reconstruction-based methods}, such as GPTQ (\cite{GPTQ}), reconstruct block outputs by minimizing quantization error iteratively, preserving the original model behavior after quantization. \textit{Rounding-based optimizations} refine the rounding process to better preserve accuracy. Representative methods are na\"ive round-to-nearest (RTN) and TesseraQ (\cite{TesseraQ}), where an iterative and progressive adaptive rounding is proposed. In addition, methods like QUIK (\cite{QUIK}) and SpQR (\cite{SpQR}) use hybrid quantization strategies to balance model accuracy and overall compression ratio. Table \ref{tab:ptq_list} summarizes the eight evaluated quantization algorithms.  

\begin{table}[h]
\centering
\begin{small}
\captionsetup{font=footnotesize}
\caption{A summary of the evaluated post-training quantization algorithms, classified in three primary categories. $X$ and $W$ represent for activations and weights, respectively. $s_j$ is the per-channel scaling factor in transformation-based methods. $H$ is second order Hessian matrix, and $E$ denotes quantization errors calculated with $H$. $\mathbb{L}$ is the block reconstruction loss function. $\gamma$ and $\beta$ are learnable parameters for weight clipping.}
\label{tab:ptq_list}
\scalebox{0.8}{
\small\begin{tabular}{c c c}
  \toprule
  Category & Algorithm & Strategy \\
  \midrule
   \multirow{3}{*}{Scaling-based Transformation} & SmoothQuant & $s_j=max\left(|X_j|\right)^\alpha/max\left(|W_j|\right)^{1-\alpha}, \alpha=0.8$ \\
  \cmidrule(r){2-3}
 & AWQ & $s_j=max\left(|X_j|\right)^\alpha/max\left(|W_j|\right)^{1-\alpha}, \text{grid search for } \alpha \in [0,1]$ \\ 
 \cmidrule(r){2-3}
 & OmniQuant & $s=\operatorname*{argmin}_s \mathbb{L}$, and $\Delta=\frac{\gamma max(W)-\beta min(W)}{2^N-1}, z=-\lfloor \frac{\beta min(W)}{\Delta}\rceil$\\
  \midrule
  Reconstruction-based & GPTQ & $W \leftarrow W - EH^{-1}, H^{-1}=\left(2XX^T+\lambda \mathbb{I}\right)^{-1} $\\
  \midrule
  \multirow{2}{*}{Rounding Optimization} & RTN & $w_q=\text{clip}\left(\lfloor\frac{w_{f}}{\Delta}\rceil + z, N_{min}, N_{max}\right)$\\
 \cmidrule(r){2-3}
 & TesseraQ & $\alpha_j = \min_{\alpha_j} \mathbb{L}, w_q=\text{clip}\left(\lfloor\frac{w_{f}}{\Delta}\rceil + \alpha_j + z, N_{min}, N_{max}\right), \alpha_j \in \{0,1\}^d $ \\
  \midrule
Hybrid & QUIK, SpQR& $W=\begin{cases} \text{lower bit-width} , & \text{non-outlier}\\
    16\text{-bit}, & \text{outlier} ($<1\%)$.
  \end{cases}$ \\
  \bottomrule
  \end{tabular}}
\end{small}
\end{table}

\subsection{Edge-ASR Models}
To evaluate the impact of post-training quantization techniques on edge-ASR performance, we select two state-of-the-art open-source ASR model families: OpenAI's Whisper (\cite{Whisper}) and Useful Sensors' Moonshine (\cite{Moonshine}). Both model families lead the open ASR leader-board across seven diverse datasets and feature compact model sizes ($27$M to $244$M parameters), making them well-suited for deployment on low-power, always-on edge devices.

\textbf{Whisper} adopts an encoder-decoder Transformer architecture. Input audio is resampled to $16$kHz, segmented into $30$-second chunks, and converted into $80$ channel log-Mel spectrograms for the encoder. The decoder generates text, intermixed with special tokens for tasks such as language identification, multilingual speech transcription, and English translation. Whisper's large-scale pretraining yields strong robustness in real-world scenarios. 

\textbf{Moonshine} follows a similar encoder-decoder Transformer structure, but is explicitly optimized for low-latency on-device inference. It takes raw audio input and incorporates rotary positional embedding (RoPE), allowing support for variable-length audio inputs. Compared to Whisper, Moonshine achieves lower latency on short segments while maintaining competitive accuracy, making it an attractive choice for constrained edge environments.
  
Together, Whisper and Moonshine represent significant advances in edge-ASR models, addressing accuracy, model capacity and latency challenges. Detailed architecture specifications for each model family are provided in Appendix \ref{appendix:architectures}. 

\subsection{Benchmarks and Datasets}
Following the open ASR leaderboard, we benchmark eight PTQ methods on two edge-ASR model families across seven evaluation datasets: AMI corpus (\cite{AMI}), Earnings-22 (\cite{Earnings-22}), GigaSpeech (\cite{GigaSpeech}), LibriSpeech (\cite{Librispeech}), SPGISpeech (\cite{SPGISpeech}), TED-Lium (\cite{TED-LIUM}) and VoxPopuli (\cite{VoxPopuli}). These datasets cover a diverse range of domains, accents and acoustic conditions, providing a robust performance baseline. 

For each quantization method, we explore multiple configurations: varying weight and activation bit-widths and quantization granularity. To align with typical support in low-power inference accelerators, we evaluate \textit{per-channel} and \textit{per-group} quantization for weights, and \textit{per-tensor} and \textit{per-token} quantization for activations. Our evaluation reports a full spectrum of deployment-relevant metrics, including word-error-rate (WER), model size, memory I/O and BOPs. The comparisons offer end-to-end insights into the trade-offs between accuracy and efficiency under different quantization strategies.  

\subsection{Statistical Analysis}
To better understand the source of quantization degradation in specific layers or configurations, we incorporate a statistical analysis based on \textit{Kurtosis} value of a tensor, defined as: 
\begin{equation}\label{eq:kurtosis}
K = \frac{1}{m}\sum_{i=1}^{m}\left(\frac{x_i-\mu}{\sigma}\right)^4\\
\end{equation}
where $m$ is the number of data points, $\mu$ is the mean and $\sigma$ is the standard deviation. Intuitively, the kurtosis value quantifies the outlier condition of a certain distribution. A higher kurtosis indicates a distribution with heavy tails and outliers, which may result in larger quantization errors. Conversely, a low kurtosis suggests a distribution with fewer outliers and typically lower quantization loss. By analyzing layer-wise kurtosis for both weights and activations, we can identify which components are most vulnerable to quantization and may benefit from techniques such as mixed-precision.

\section{Benchmarking Quantized edge-ASR Models}
\label{benchmarking}
This section presents a comprehensive benchmarking of PTQ methods on edge-ASR models. We begin with the experimental setup (subsection \ref{sub:experiment}), followed by an in-depth quantization result analysis (subsection \ref{sub:quant_analysis}), covering bit-width selection, ultra-low bit quantization, granularity, weight clipping, and the use of symmetric versus asymmetric quantization. We also compare deployment relevant metrics, including WER, model size, memory I/O and BOPs in subsection \ref{sub:fullcomparison}. Additional ablation studies on the impact of calibration data source and size are further discussed in Appendix \ref{appendix:ablation}.

\subsection{Experimental Setup}
\label{sub:experiment}
We outline the experimental setup used to evaluate quantized edge-ASR models. Additional implementation details are available in Appendix \ref{appendix:architectures} - \ref{appendix:dataset}.

\textbf{Models.} We assess two ASR families optimized for edge deployment. Three variants: Tiny ($39$M), Base ($74$M) and Small ($244$M), are selected from the Whisper family, while the Moonshine family includes Tiny ($27$M) and Base ($61$M). Both model families share an encoder-decoder Transformer architecture but differ in positional embedding and other configurations. We employ a byte-level BPE tokenizer (vocabularies of $51,864$ for Whisper and $32,768$ for Moonshine), and apply standard text normalization before computing WER.    

\textbf{Datasets.} Following the open ASR leader-board, we evaluate on seven benchmark datasets. These datasets cover diverse domains, accents and speaking styles, with audio durations ranging from $5$ to $100$ hours. Our floating-point baselines match reported leader-board results (\cite{Leaderboard}). For algorithms that require calibration, we use a fixed set of $256$ English utterances sampled from the Mozilla Common Voice (\cite{CommonVoice}) that are different from any evaluation dataset, ensuring a fair comparison. Table \ref{tab:benchmark_list} summarizes the full benchmark setup.

\begin{table}[!t]
\centering
\begin{small}
\captionsetup{font=footnotesize}
\caption{Experimental setup for benchmarking quantized edge-ASR models.}
\label{tab:benchmark_list}
 \scalebox{0.8}{
  \begin{tabular}{c | c }
  \toprule
  PTQ algorithms &  SmoothQuant, AWQ, OmniQuant, GPTQ, RTN, TesseraQ, QUIK, SpQR\\
  \midrule
 \multirow{2}{*}{ granularities} & weights: per-channel, per-group\\
& activations: per-tensor, per-token \\
  \midrule
  edge-ASR models & Whisper (Tiny, Base, Small), Moonshine (Tiny, Base) \\
  \midrule
  Datasets &  AMI corpus, Earnings-22, GigaSpeech, LibriSpeech (clean, other), SPGISpeech, TED-Lium,  VoxPopuli\\
  \midrule
  Evaluation metrics & WER, kurtosis value, model size, memory I/O, bit operations (BOPs)\\
  \bottomrule
  \end{tabular}}
\end{small}
\end{table}

\subsection{Quantization Analysis}
\label{sub:quant_analysis}

\subsubsection{Bit-width selection}
\label{sub:bitwidth}
To determine the optimal bit-width for ASR models, we evaluate WER under various PTQ algorithms across all benchmark datasets. Table \ref{tab:wer} reports the averaged WER of the $5$ quantized edge-ASR models across $7$ datasets, alongside its floating-point baseline. Detailed per-dataset WER performances of each model are provided in Appendix \ref{appendix:quantperformance}. 

Our results show that PTQ with $8$-bit weights and $8$-bit activations (w8-a8) reliably preserves model performance across all $5$ models and quantization algorithms. Within both model families, reducing weight precision to $4$ bits leads to notable degradation of performance metrics in smaller models such as Whisper Tiny (Table \ref{tab:whisper_tiny}) and Moonshine Tiny (Table \ref{tab:moonshine_tiny}), while larger models, including Whisper Base (Table \ref{tab:whisper_base}), Whisper Small (Table \ref{tab:whisper_small}) and Moonshine Base (Table \ref{tab:moonshine_base}) exhibit strong robustness under the same low-bit quantization. For the Moonshine family, SmoothQuant performs poorly at low weight bit-width, and only becomes effective when weight precision increases to $8$-bit. 

\begin{table}[h]
\centering
\begin{small}
\captionsetup{font=footnotesize}
\caption{Overall performance of quantized edge-ASR models under diverse bit-width configurations across $8$ quantization algorithms. The averaged WER across 7 datasets is reported. Detailed per-dataset WER performances of each model are provided in Appendix \ref{appendix:quantperformance}. To align with typical hardware support, per-tensor symmetric quantization is used for activations, and per-group symmetric quantization is used for weights, unless otherwise specified. The group size of weights for Whisper models, Moonshine Tiny and Moonshine Base are $64$, $72$ and $52$, respectively. The underlined value shows the best performance among $8$ quantization algorithms within the same bit-width configuration.}
\label{tab:wer}
 \scalebox{0.8}{
  \begin{tabular}{c c c c c c c}
  \toprule
  \multirow{2}{*}{Bits}  &  \multirow{2}{*}{Method} & \multicolumn{5}{c}{\textbf{averaged WER \%} $\mathbf{\downarrow}$} \\
  \cmidrule(r){3-7}
  &  & Whisper Tiny & Whisper Base & Whisper Small & Moonshine Tiny & Moonshine Base\\
  \midrule
  Float &  Baseline & 12.80 &10.32 &8.59 &12.72 &9.99 \\
  \midrule
  \multirow{8}{*}{w4-a8}  &	SmoothQuant & 31.78 &16.81& 9.68&72.99&77.22\\
  & AWQ & 						         21.52&	11.87& 8.82&15.32&10.98\\
  & OmniQuant & 						   20.31&	11.77&8.64&14.74&10.74\\
  & GPTQ & 	      						   18.57&	11.43&8.74&\uline{14.15}&10.38\\
  & RTN &  						         21.30&	11.44&8.78&25.38&13.48\\
  & TesseraQ & 						  17.05&	11.40&8.73&17.95&12.08\\
  & QUIK & 							 \uline{14.69}&\uline{11.06}&8.86&14.32&10.37\\
  & SpQR & 						       15.75&11.14&\uline{8.60}&\uline{14.15}&\uline{10.31}\\
  \midrule
  \multirow{8}{*}{w4-a16}  & SmoothQuant & 21.90&12.49&9.00&69.81&14.20\\
  & AWQ &   							   17.83&	11.41&8.92&15.12&10.57\\
  & OmniQuant &  						  16.85&	11.59&8.92&14.47&10.54\\
  & GPTQ & 							16.46&11.16&8.77&\uline{13.61}&10.28\\
  & RTN & 							17.78&11.36&8.86&23.23&12.94\\
  & TesseraQ &  						17.58&11.42&8.87&15.21&10.78\\ 
  & QUIK &  							14.73&11.30&8.90&14.24&10.40\\
  & SpQR &  						     \uline{14.24}&\uline{10.94}&\uline{8.65}&13.81&\uline{10.14}\\
  \midrule
  \multirow{8}{*}{w8-a8}  & SmoothQuant &   13.93&10.82&8.55&13.42&38.87\\
  & AWQ &  								14.04&10.40&8.64 &13.12&10.16\\ 
  & OmniQuant & 						      13.86&11.26&\uline{8.44}&13.04&10.14\\
  & GPTQ &   							      13.65&10.52&8.45&13.09&10.15\\
  & RTN &   								13.39&10.51&8.56&13.05&10.09\\
  & TesseraQ &  							13.16&10.59&8.47&13.50&10.29\\
  & QUIK & 								\uline{12.94}&\uline{10.18}&8.63&\uline{12.81}&\uline{10.04}\\
  & SpQR & 								14.04&10.54&8.54&13.04&10.12\\
  \midrule
  \multirow{8}{*}{w8-a16}  & SmoothQuant & 12.76&10.49&8.62&12.78&\uline{9.98}\\
  & AWQ &  							12.85&10.21&\uline{8.49}&12.78&10.06\\
  & OmniQuant & 						      13.09&10.41&8.58&\uline{12.73}&10.05\\
  & GPTQ &   							      12.91&10.40&8.61&12.77&9.99\\
  & RTN &   								   12.78&	10.30&8.64&12.76&10.03\\
  & TesseraQ &  							\uline{12.68}&\uline{10.17}&8.64&12.74&10.05\\
  & QUIK & 									12.72&10.28&8.62&12.74&10.04\\
  & SpQR & 									12.73&10.25 &8.59&12.76&10.03\\ 
  \bottomrule
  \end{tabular}}
\end{small}
\end{table}

Interestingly, the w8-a8 configuration consistently surpasses w4-a16, highlighting the importance of weight quantization. Overall, reconstruction-based and hybrid quantization methods outperform scale-based transformations under low-bit settings. These patterns are consistent across all Whisper and Moonshine variants, confirming that while $8$-bit precision is generally safe for deployment, lower-bit weight quantization requires careful algorithmic design and model capacity considerations. As a reference, the runtime cost of each quantization algorithm is available in Appendix \ref{appendix:runtime}.

\subsubsection{Ultra-low bit quantization}
\label{sub:lowbit}
To explore the limits of PTQ under extreme compression, we evaluate ultra-low bit weight quantization on the Whisper Base and Moonshine Base models. In these experiments, weights are quantized asymmetrically, and activations use per-token quantization to maximize their representation power. 


\begin{table}[!t]
\centering
\begin{small}
\captionsetup{font=footnotesize}
\caption{Ultra-low bit quantization performance for Whisper Base and Moonshine Base models across $7$ datasets under $8$ quantization algorithms. Per-token symmetric quantization is used for activations, and per-group asymmetric quantization is used for weights. The group sizes for Whisper and Moonshine are $64$ and $52$, respectively. The underlined value shows the best quantization performance in each dataset, and the best averaged WER across all datasets is highlighted.}
\label{tab:ultralowbit}
\begin{adjustbox}{width=\textwidth}
  \begin{tabular}{c c c c c c c c c c c}
  \toprule
  \multirow{2}{*}{Model}  &  \multirow{2}{*}{Method} & \multicolumn{9}{c}{\textbf{w3-a16 (WER \%)} $\mathbf{\downarrow}$} \\
  \cmidrule(r){3-11}
  &  & AMI & Earnings-22 & GigaSpeech & Libri clean & Libri other & SPGISpeech & TED-Lium & Voxpopuli & Avg.\\
  \midrule
  \multirow{9}{*}{Whisper Base}  &  Float &  21.13 & 15.09 & 12.87 & 4.28 &	10.36 & 4.27 & 4.85 & 9.75 & 10.32 \\
\cmidrule(r){2-11}
  & SmoothQuant & 100.94&	207.28&	113.70&	199.29&	185.28&	128.14&	144.16&	216.92&		161.96\\
  & AWQ & 								31.10&	36.80&	19.71&	9.96&	21.65&	11.15&	9.58&	22.29&	20.28     \\
  & OmniQuant & 						     30.79&	26.64&	17.85&	7.91&	15.24&	8.55&	7.70&	15.74&		16.30\\
  & GPTQ & 	      							     36.65&	35.24&	24.93&	11.38&	22.27&	10.98&	11.48&	22.15&		21.89\\
  & RTN &  								     27.23&	29.85&	17.49&	9.19&	15.27&	8.78&	7.38&	16.28&		16.43\\
  & TesseraQ & 								28.20&	27.36&	16.88&	8.32&	16.81&	8.64&	7.13&	16.05&		16.17\\
  & QUIK & 								    23.16&	20.05&	\uline{14.28}&	6.36&	13.67&	5.95&	\uline{5.63}&	14.18&		12.91 \\
  & SpQR & 								    \uline{22.64}&	\uline{18.36}&	14.44&	\uline{5.01}&	\uline{12.73}&	\uline{5.18}&	6.50&	\uline{12.34}&		\textbf{12.15} \\
 \midrule
 \multirow{9}{*}{Moonshine Base}  &	Float &  17.07&	17.69&	12.11&	3.26&	8.28&	5.46&	5.24&	10.79&		9.99\\
\cmidrule(r){2-11}
  & SmoothQuant & 33.95&	36.94&	21.78&	13.74&	29.93&	16.69&	12.41&	22.42&		23.48\\
  & AWQ & 								20.86&	24.37&	14.12&	4.24&	11.25&	7.05&	6.49&	12.36&	12.59     \\
  & OmniQuant & 						     21.21&	22.85&	13.91&	4.22&	10.72&	7.34&	6.41&	12.90&		12.45\\
  & GPTQ & 	      							     18.92&	20.18&	13.08&	3.92&	9.79&	6.48&	5.96&	12.10&		11.30\\
  & RTN &  								     90.01&	117.00&	74.32&	70.87&	101.09&	89.60&	73.99&	115.73&		91.58\\
  & TesseraQ & 								93.03&	122.89&	76.53&	70.73&	105.21&	94.94&	77.54&	119.75&		95.08\\
  & QUIK & 								    20.44&	20.12&	13.21&	3.95&	10.23&	6.95&	6.01&	11.72&		11.58 \\
  & SpQR & 								    \uline{18.45}&	\uline{19.33}&	\uline{12.49}&	\uline{3.75}&	\uline{9.11}&	\uline{5.89}&	\uline{5.13}&	\uline{11.23}&		\textbf{10.67} \\
\bottomrule
 \end{tabular}
\end{adjustbox}
\end{small}
\end{table}

Enforcing $2$-bit weight precision causes all PTQ methods to fail, resulting in high WER. For weights below $3$-bit, Quantization-Aware Training (QAT) becomes essential. However, at $3$-bit weight precision, several algorithms still produce promising results. As shown in Table \ref{tab:ultralowbit}, OmniQuant and hybrid strategies (e.g., QUIK, SpQR that preserves higher precision for the top $1\%$ outliers) maintain reasonable accuracy. In contrast, scaling-based transformation such as SmoothQuant perform poorly on the Whisper Base model. Similarly, RTN and TesseraQ struggle on the Moonshine Base model at this precision, indicating that their rounding-based strategies alone are insufficient at ultra-low precision. %

\begin{figure}[!b]
  \centering
  \vspace*{-4mm}
  \includegraphics[width=1.0\linewidth]{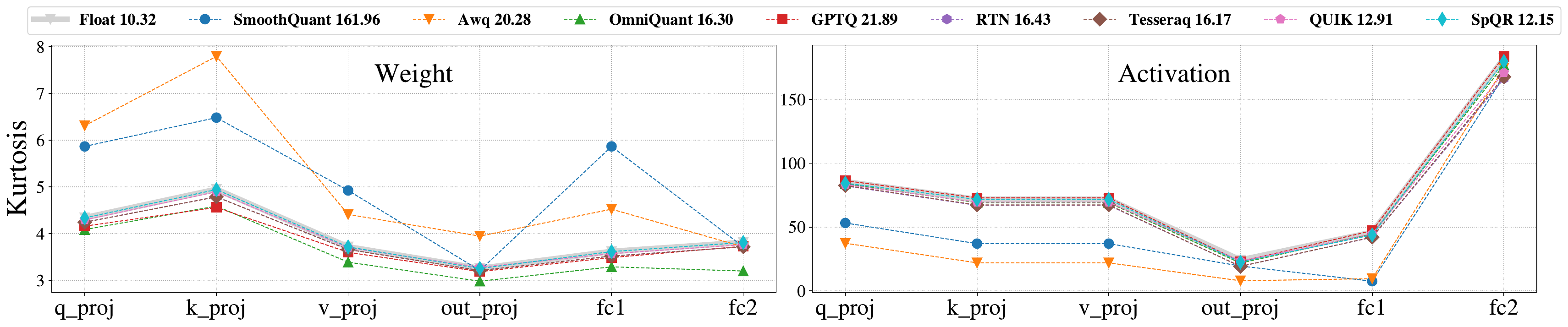}
  \caption{Averaged kurtosis value of weights and input activations with various layer types in Whisper Base for different methods under w3-a16 quantization. The legends denote the quantization method and its corresponding averaged WER across $7$ datasets.}
  \label{fig:kurtosis}
\end{figure}

To further examine each algorithm's internal behavior, we analyze the Whisper Base model and present the averaged kurtosis value of weights and activations over all layers in Figure \ref{fig:kurtosis}. This provides a statistical perspective on the distribution characteristics after quantization. We observed that scaling-based transformation methods (SmoothQuant, AWQ) achieve lower kurtosis for activations at the cost of increased weight kurtosis, due to the design principle of offloading the quantization difficulty from activations to weights. This justifies the significant performance degradation under ultra-low bit settings, even when activations remain at $16$ bits. These results underscore the need for algorithmic robustness and outlier handling when pushing PTQ to its limits. 

\subsubsection{Quantization granularity}
\label{sub:granularity}
To assess the impact of weight and activation quantization granularity, we select three representative PTQ algorithms (AWQ, GPTQ and TesseraQ), one from each category, and apply $4$-bit uniform quantization to weights under two schemes: per-channel and per-group. $16$-bit per-tensor and per-token quantization are applied on activations. Table \ref{tab:granularity} reports the averaged WER across all seven datasets for each model and granularity setting.

\begin{table}[h]
\centering
\vspace*{-2mm}
\begin{small}
\captionsetup{font=footnotesize}
\caption{The impact of weight and activation quantization granularity to model performances. The averaged WER across $7$ datasets is reported. Detailed per-dataset WER performances are available in Appendix \ref{appendix:granularity}. For per-group weight quantization, the coarse group sizes for Whisper, Moonshine Tiny and Moonshine Base are $128$, $144$ and $104$, and the fine group sizes are $64$, $72$ and $52$, respectively. The underlined value shows the best quantization performance in each category.}
\label{tab:granularity}
 \scalebox{0.8}{
  \begin{tabular}{c c c c c c c c}
  \toprule
  \multirow{2}{*}{Method}  & \multicolumn{2}{c}{granularity} &  \multicolumn{5}{c}{\textbf{w4-a16 Models (avg. WER \%)} $\mathbf{\downarrow}$} \\
 \cmidrule(r){2-3}
  \cmidrule(r){4-8}
   & weights & activations & Whisper tiny & Whisper base & Whisper small & Moonshine tiny & Moonshine base \\
  \midrule
  Float & & & 12.80 & 10.32 & 8.59 & 12.72 & 9.99 \\
  \midrule
  AWQ & \multirow{3}{*}{per-channel} &  \multirow{3}{*}{per-tensor}  & 64.34 & 85.64 & 9.94 & 15.67 & 11.16\\
  GPTQ & & & \uline{28.04} & \uline{15.32} & \uline{9.27} & \uline{14.60} & \uline{10.74}\\
  TesseraQ & & & 43.58 & 18.06 & 9.42 & 16.26 & 11.43\\
\midrule
   AWQ & \multirow{3}{*}{per-group, coarse} & \multirow{3}{*}{per-tensor} &  23.26 & 12.64 & 8.86 & 15.17 & 10.79 \\
  GPTQ&  & & \uline{16.84} & \uline{11.60} & \uline{8.81} & \uline{13.91} & \uline{10.50}\\
  TesseraQ&  & &  20.51 & 11.75 & 8.96 & 15.65 & 11.20\\ 
  \midrule
   AWQ & \multirow{3}{*}{per-group, fine} & \multirow{3}{*}{per-tensor} & 17.83 & 11.41 & 8.92 & 15.12& 10.57 \\
  GPTQ&  & &\uline{16.46} & \uline{11.16} & \uline{8.77} & \uline{13.61} &  \uline{10.28}\\
  TesseraQ&  & & 17.58 & 11.42 & 8.87 & 15.21 & 10.78 \\  %
 \midrule
   AWQ & \multirow{3}{*}{per-group, fine} & \multirow{3}{*}{per-token} & 17.82 & 11.38 & 8.96 & 14.95& 10.64 \\
  GPTQ&  & &\uline{16.72} & 11.88 & 8.95 & \uline{13.67} &  \uline{10.30}\\
  TesseraQ&  & & 17.78 & \uline{11.36} & \uline{8.86} & 15.24 & 10.77 \\  %
\bottomrule
 \end{tabular}}
\end{small}
\end{table}

Our results show that GPTQ consistently delivers the lowest averaged WER across all weight granularity configurations, demonstrating the strong robustness to coarse quantization. For larger models (Moonshine Base and Whisper Small), per-channel weight quantization shows reasonable accuracy, making it an efficient option. In contrast, smaller models (Whisper Tiny and Whisper Base) benefit notably from per-group quantization, where fine-grained scaling helps mitigate quantization noise and significantly improves WER. We also observe that activation granularity has minimal impact at $16$-bit precision. These findings highlight the importance of choosing the appropriate weight quantization granularity, especially when targeting low capacity models on hardware with limited quantization support.

\subsubsection{Weight clipping}
\label{sub:clipping}
The technique of weight clipping is commonly used to constrain the range of weight values prior to quantization. By suppressing extreme outliers, it reduces quantization error. However, this process may also introduce clipping loss, potentially harming model accuracy. To investigate this trade-off, we use AWQ as a case study to evaluate the impact of weight clipping under four bit-width settings: w4-a8, w4-a16, w8-a8 and w8-a16 on both Whisper Tiny and Whisper Base models. Table \ref{tab:weigt_clipping} reports the resulting average WER across all seven evaluation datasets. 

\begin{table}[h]
\centering
\vspace*{-2mm}
\begin{small}
\captionsetup{font=footnotesize}
\caption{Impact of weight clipping in AWQ to Whisper Tiny and Base models under various bit-width settings. Averaged WER across $7$ datasets is reported. Associated per-dataset WER performances are shown in Appendix \ref{appendix:weightclipping}. Per-tensor quantization is used for activations, and per-group quantization is used for weights, with a group size of $64$. Both weights and activations are using symmetric quantization. The underlined value displays the best quantization performance in each category.}
\label{tab:weigt_clipping}
\scalebox{0.9}{
  \begin{tabular}{c c c c c c c c c}
\toprule
  \multirow{2}{*}{\textbf{avg. WER \% $\mathbf{\downarrow}$}}  & \multicolumn{2}{c}{w4-a8} &  \multicolumn{2}{c}{w4-a16} &  \multicolumn{2}{c}{w8-a8} &  \multicolumn{2}{c}{w8-a16}\\
\cmidrule(r){2-3}
\cmidrule(r){4-5}
\cmidrule(r){6-7}
\cmidrule(r){8-9}
 & w/ clip & w/o clip & w/ clip & w/o clip & w/ clip & w/o clip & w/ clip & w/o clip \\
  \midrule
  Whisper Tiny & 43.67 & \uline{21.52} & 42.53 & \uline{17.83} & 16.76 & \uline{14.04} & 14.80 & \uline{12.85}\\
  \midrule
 Whisper Base & 13.30 & \uline{11.87} & 13.27 & \uline{11.41} & \uline{10.40} & 15.96 & \uline{10.21} & 10.32\\
  \bottomrule
  \end{tabular}}
\end{small}
\end{table}

Our experiments reveal a clear capacity-dependent effect. For Whisper Tiny model, weight clipping consistently degrades performance across all bit-width settings. The clipping error introduced by capping weight values outweighs any reduction in quantization noise, leading to higher WER. For Whisper Base model, weight clipping demonstrates measurable benefits as precision increases. At $8$-bit weight settings, clipped models achieve modest WER reductions compared to their unclipped counterparts. These findings indicate that weight clipping should be applied selectively. It can enhance quantization robustness for higher capacity models operating at higher precisions, but tend to harm smaller models, which are less resilient to additional clipping distortion.

\subsubsection{Symmetric or asymmetric}
\label{sub:asymmetric}
Symmetric quantization with a zero offset is generally more hardware-friendly and thus preferred for deployment. However, asymmetric quantization introduces an additional grid point in the quantized space, which can improve model accuracy, especially under low bit-width settings. To quantify this effect, we compare symmetric and asymmetric weight quantization across all eight PTQ algorithms on Whisper Base and Moonshine Base models. As shown in Table \ref{tab:asymmetric}, asymmetric quantization outperforms symmetric quantization in $7$ out of $8$ algorithms. Notably, for algorithms like SmoothQuant and AWQ, asymmetric quantization proves to be crucial to achieving acceptable WER when using $4$-bit per-channel weight quantization. These results highlight that, despite its slightly higher implementation complexity, asymmetric quantization often yields superior performance, making it a valuable strategy in low-bit deployment scenarios. 

\begin{table}[h]
\centering
\captionsetup{font=footnotesize}
\caption{Symmetric vs Asymmetric weight quantization. The averaged WER across $7$ datasets is reported. Associated per-dataset WER performances are available in Appendix \ref{appendix:asymmetric}. $4$-bit per-channel and per-group weight quantizations are evaluated. To eliminate other impacts, activations are fixed with $16$-bit per-token quantization. Performance benefits from asymmetric quantization that are larger than $1\%$ are underlined.}
\label{tab:asymmetric}
\scalebox{0.7}{
  \begin{tabular}{c c c c c c c c c}
  \toprule
 \multirow{3}{*}{Method}   & \multicolumn{4}{c}{\textbf{Whisper Base (WER \%) $\mathbf{\downarrow}$}} & \multicolumn{4}{c}{\textbf{Moonshine Base (WER \%) $\mathbf{\downarrow}$}} \\
\cmidrule(r){2-5}
\cmidrule(r){6-9}
 &  \multicolumn{2}{c}{per-group, g=64} & \multicolumn{2}{c}{per-channel} & \multicolumn{2}{c}{per-group, g=52} & \multicolumn{2}{c}{per-channel} \\
\cmidrule(r){2-3}
\cmidrule(r){4-5}
\cmidrule(r){6-7}
\cmidrule(r){8-9}
 & asym. & sym. & asym. & sym. & asym. & sym. & asym. & sym. \\
\midrule
SmoothQuant & 11.53 & 12.44 & \uline{41.64} & 616,57 & 11.40 & 14.20 & \uline{57.98} & 99.67\\
AWQ & 10.95 & 11.38 & \uline{14.12} & 50.65 & 10.39 & 10.64 & 10.75 & 11.13 \\
OmniQuant & 11.67 & 11.61 & 18.38 & 16.22 & 10.41 & 10.39 & 11.17 & 11.07 \\
GPTQ & \uline{10.81} & 11.88 & \uline{12.62} & 15.84 & 10.19 & 10.30 & 10.53 & 10.54 \\
RTN & 11.05 & 11.42 & \uline{12.61} & 17.66 & \uline{11.18} & 12.90 & \uline{16.11} & 18.77 \\
TesseraQ & 11.14 & 11.36 & \uline{12.41} & 17.66 & \uline{11.16} & 12.97 & \uline{16.09} & 18.78\\
QUIK & 10.65 & 11.22 & 11.05 & 11.82 & 10.29 & 10.40 & 10.63 & 10.87 \\
SpQR & 10.98 & 10.93 & 11.19 & 11.14 & 10.27 & 10.15 & 10.42 & 10.51 \\
 \bottomrule
  \end{tabular}}
\end{table}

\subsection{Overall Comparison} 
\label{sub:fullcomparison}

Table \ref{tab:comparison} showcases the overall comparisons of quantized edge-ASR models across key deployment-relevant metrics. \textit{Model size} refers primarily to the total memory required to store model parameters. This has a significant impact on both latency and power consumption, particularly when weights cannot fit on-chip, resulting in frequent external memory accesses. \textit{Memory I/O} captures the data movement involved in loading inputs and weights, and storing outputs for each computational operation. Since memory I/O can dominate both latency and energy usage, minimizing memory access is critical for achieving energy-efficient and low-latency inference. \textit{Bit Operations (BOPs)} measures the total computational cost, taking into account both the number of MAC (multiply-accumulate) operations and data bit-width. Higher BOPs correspond to increased computing energy. Therefore, reducing bit-width not only lowers model size, but also significantly decreases the energy consumed per operation. For ASR models, the encoder is constrained by activation size, where memory I/O dominates the energy usage, while the decoder is constrained by weight size. Therefore, for low-power use cases, configurations such as w8-a8 for encoder, and w4-a16 for decoder would be considered for edge application. Based on the accuracy criteria and hardware specifications, users can choose appropriate models and bit-width configurations upon deployment.

\begin{table}[h!]
\centering
\begin{small}
\captionsetup{font=footnotesize}
\caption{Overall comparisons of quantized edge-ASR models across key deployment-relevant metrics. Calculation is based on $30$ second audio input. Relative GBOPs to floating-point baseline is computed. The best averaged WER across $8$ evaluated algorithms is reported.}
\label{tab:comparison}
\scalebox{0.7}{
  \begin{tabular}{c c c c c c c c}
  \toprule
 \multirow{2}{*}{Models} & \multirow{2}{*}{\# bits w/a} &  \multicolumn{2}{c}{Encoder} & \multicolumn{2}{c}{Decoder} & \multirow{2}{*}{Rel. GBOPs (\%)}  & \multirow{2}{*}{ ave. WER (\%)} \\
\cmidrule(r){3-4}
\cmidrule(r){5-6}
 & &  weight size (MB) & memory I/O (MB) & weight size (MB) & memory I/O (MB) &  &\\
  \midrule
 \multirow{5}{*}{ Whisper tiny} & 32/32 & 30.53 & 1068.35 & 118.21 & 161.55 & 100 & 12.80 \\
 & 4/8 & 3.82 & 263.27 & 14.78 & 25.61 & 3.13 & 14.69\\
 & 4/16 & 3.82 & 522.72 & 14.78 & 36.45 & 6.25 &14.24\\
 & 8/8 & 7.63 & 267.09 & 29.55 & 40.39 & 6.25 &12.94 \\
 & 8/16 & 7.63 & 526.54 & 29.55 & 51.22 & 12.50 &12.68\\
 \midrule
 \multirow{5}{*}{ Whisper base} & 32/32 & 79.29 & 2143.21 & 208.01 & 294.69 & 100 & 10.32\\
 & 4/8 & 9.91 & 525.89 & 26.00 & 47.67 & 3.13 & 11.06\\
 & 4/16 & 9.91 & 1041.87 & 26.00 & 69.34 & 6.25 & 10.94\\
 & 8/8 & 19.82 & 535.80 & 52.00 & 73.67 & 6.25 & 10.18\\
 & 8/16 & 19.82 & 1051.78 & 52.00 & 95.34 & 12.50 & 10.17\\
 \midrule
 \multirow{5}{*}{ Whisper small} & 32/32 & 348.01 & 6505.61 & 614.32 & 874.36 & 100 & 8.59\\
 & 4/8 & 43.50 & 1582.90 & 76.79 & 141.80 & 3.13 &8.60\\
 & 4/16 & 43.50 & 3122.30 & 76.79 & 206.81 & 6.25 &8.65\\
 & 8/8 & 87.00 & 1626.40 & 153.58 & 218.59 & 6.25 &8.44\\
 & 8/16 & 87.00 & 3165.80 & 153.58 & 283.60 & 12.50 &8.49\\
 \midrule
 \multirow{5}{*}{ Moonshine tiny} & 32/32 & 30.72 & 195.23 & 77.65 & 95.14 & 100 & 12.72\\
 & 4/8 & 3.84 & 44.97 & 9.71 & 14.08 & 3.13 & 14.15\\
 & 4/16 &3.84 & 86.09 & 9.71 & 18.45 & 6.25 & 13.61\\
 & 8/8 & 7.68 & 48.81 & 19.41 & 23.78 & 6.25 & 12.81\\
 & 8/16 & 7.68 & 89.93 & 19.41 & 28.16 & 12.50 &12.73\\
 \midrule
 \multirow{5}{*}{ Moonshine base} & 32/32 & 80.61 &371.40 & 165.44 & 199.13 & 100 & 9.99\\
 & 4/8 & 10.08 & 82.77 & 20.68 & 29.10 & 3.13 & 10.31\\
 & 4/16 & 10.08 & 155.47 & 20.68 & 37.52 & 6.25 & 10.14\\
 & 8/8 & 20.15 & 92.85 & 41.36 & 49.78 & 6.25 & 10.04 \\
 & 8/16 & 20.15 & 165.54 & 41.36 & 58.20 & 12.50 & 9.98\\
  \bottomrule
  \end{tabular}}
\end{small}
\end{table}

\section{Limitations}
\label{limitations}

While our benchmark provides a comprehensive assessment of post-training quantization for edge-ASR models, several limitations remain. \textit{Model diversity.} Our study focuses on two transformer-based ASR model families, Whisper and Moonshine, due to their compact model sizes, leading performance, and suitability for edge deployment. Future work will expand the edge-ASR models to include additional architectures (e.g., RNN-based, state space models (SSMs), Conformer variants and Canary), to broaden the applicability of our findings. \textit{Algorithm coverage.} We evaluate eight state-of-the-art PTQ methods covering transformation, reconstruction and rounding-based strategies. Emerging techniques, such as rotation-based quantization, are not yet supported in our current toolkit. Incorporating these newer algorithms will enrich the benchmark and offer more effective options for aggressive model compression. \textit{Framework extensibility.} Although we extend the LLM compression toolkit to support ASR models and a diverse set of quantization algorithms, many other factors remain unexplored (e.g., dynamic quantization, hardware-specific quantization strategies). Future enhancements will expand our framework to include these dimensions, enabling comprehensive support for end-to-end model optimization, conversion and deployment on a wide array of edge platforms. Addressing these limitations in future work will help guide the development of more efficient and accurate quantized ASR systems tailored for resource-constrained environments. 

\section{Conclusion}
\label{conclusion}

We present the first systematic benchmark of eight post-training quantization (PTQ) methods on two Transformer-based edge-ASR model families (Whisper and Moonshine), evaluated across seven diverse datasets. Through comprehensive analysis of critical quantization factors, including bit-width selection, ultra-low bit quantization, granularity, weight clipping, and symmetric versus asymmetric quantization, along with ablation studies on calibration data, we characterize the trade-offs between model accuracy, model size, memory footprint, I/O overhead, and computational cost in the fixed-point domain, establishing a quantized standard for the open ASR leader-board. Our findings show that advanced PTQ strategies enable viable quantization down to $3$-bit weights. By building an extension to the LLM compression toolkit to support ASR architectures, we provide a workflow that enables rapid, reproducible quantization of edge-ASR models. These insights and tools serve to bridge the gap between model training and edge deployment. Future work will broaden the model zoo to include a wide range of ASR architectures, incorporate emerging quantization techniques, and integrate  deployment pipelines for edge platforms, facilitating efficient and accurate quantized ASR systems on low-power, always-on edge devices.


\clearpage
\printbibliography

\clearpage
\begin{appendices}
\appendix

\section{Architectures of edge-ASR Model Families} 
\label{appendix:architectures}

\begin{table}[h!]
\centering
\captionsetup{font=footnotesize}
\caption{Architecture details of the Whisper and Moonshine model families.}
\begin{center}
\label{tab:whispermoonshine}
\scalebox{0.8}{
  \begin{tabular}{c  c  c c | c  c}
  \toprule
\multirow{4}{*}{Models} & \multicolumn{3}{c}{Whisper} & \multicolumn{2}{c}{Moonshine} \\
\cmidrule(r){2-6}
  & Whisper Tiny & Whisper Base & Whisper Small & Moonshine Tiny & Moonshine Base \\
 \midrule
 Dimension & 384 & 512 & 768  &288 &  416 \\
 \midrule
 Encoder layers & 4 & 6 & 12 & 6 & 8 \\
 \midrule
 Decoder layers & 4 & 6 & 12& 6 &  8  \\
 \midrule
 Attention heads & 6 & 8 & 12 & 8 & 8 \\
 \midrule
 Encoder FFN activation & \multicolumn{5}{c}{GELU} \\
 \midrule
 Decoder FFN activation & GELU & GELU & GELU &  SwiGLU & SwiGLU \\
 \midrule
 Parameters in Millions & 37.8 & 72.6 & 244 & 27.1 & 61.5 \\
  \bottomrule
  \end{tabular}}
\end{center}
\end{table}

\section{Quantization Mechanisms} 
\label{appendix:quantschemes}

\begin{figure}[h!]
  \centering
  \includegraphics[scale=0.45]{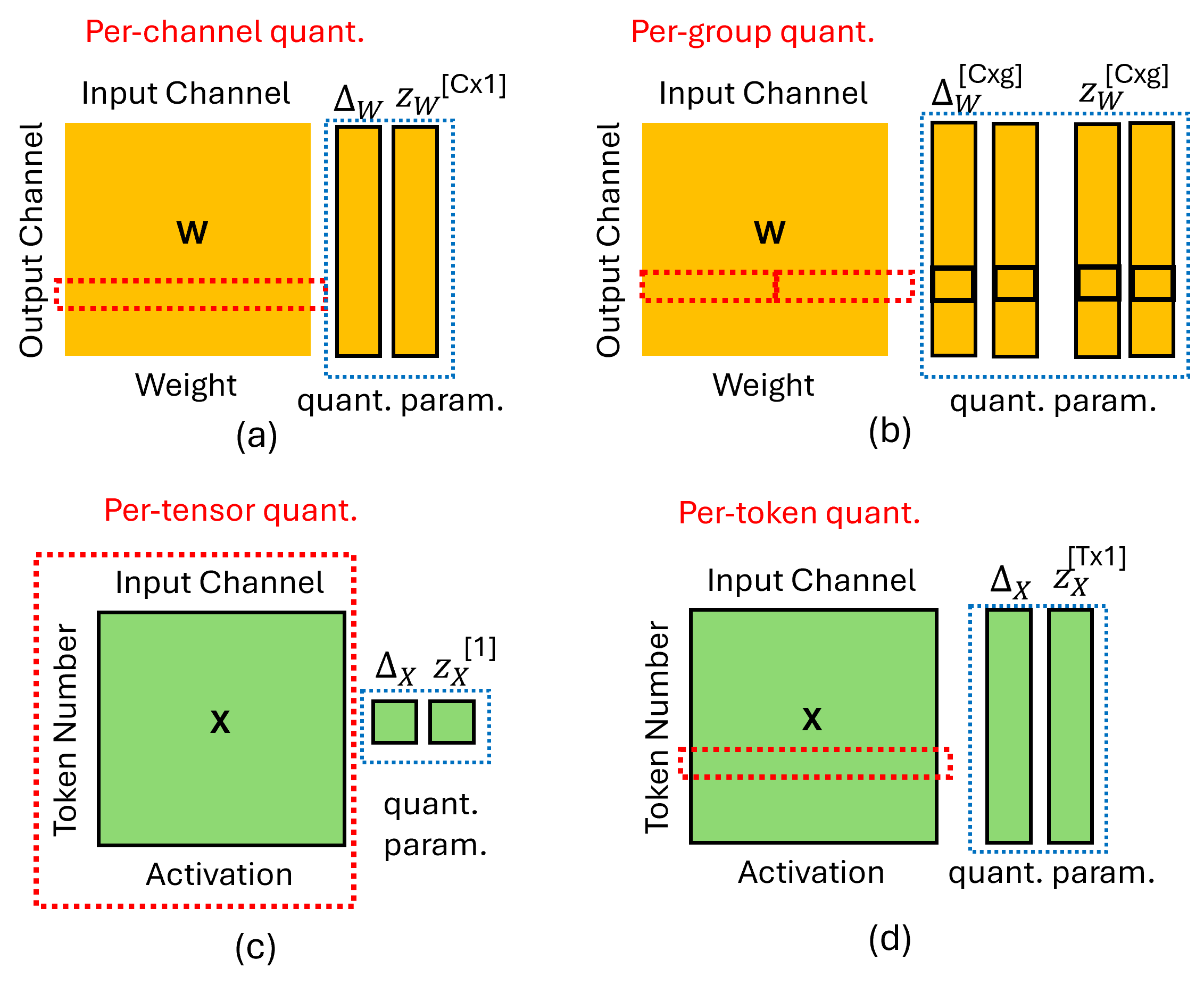} 
  \caption{Diverse quantization schemes for weights and activations. (a) Per-channel weight quantization. (b) Per-group weight quantization. (c) Per-tensor activation quantization. (d) Per-token activation quantization.}
  \label{fig:quantscheme}
\end{figure}

\section{Dataset Description} 
\label{appendix:dataset}

\begin{table}[h!]
\centering
\captionsetup{font=footnotesize}
\caption{Summary of evaluation datasets on open ASR leaderboard.}
\begin{center}
\label{tab:dataset}
\resizebox{1.0\textwidth}{!}{
  \begin{tabular}{c c c c c }
  \toprule
 Dataset & Domain & Speaking Style & Test (hours) & Transcriptions \\
 \midrule
 LibriSpeech & Audiobook & Narrated & 11 & Normalised \\
 Voxpopuli & European Parliament & Oratory & 5 & Punctuated \\
 TED-LIUM & TED talks & Oratory & 3 & Normalised \\
 GigaSpeech & Audiobook, podcast, YouTube & Narrated, spontaneous & 40 & Punctuated \\
SPGISpeech & Financial meetings & Oratory, spontaneous & 100 & Punctuated \& Cased \\
Earnings-22 & Financial meetings & Oratory, spontaneous & 5 & Punctuated \& Cased \\
AMI & Meetings & Spontaneous & 9 & Punctuated \& Cased \\
  \bottomrule
  \end{tabular}}
\end{center}
\end{table}

\section{Model Quantization Performances} 
\label{appendix:quantperformance}

\subsection{Whisper Tiny}
\begin{table}[h]
\centering
\captionsetup{font=footnotesize}
\begin{small}
\caption{Overall performance of quantized Whisper Tiny model across $7$ datasets under $8$ quantization algorithms. To align with typical hardware support, per-tensor symmetric quantization is used for activations, and per-group symmetric quantization is used for weights, unless otherwise specified. The group size for weights is $64$ for Whisper models. For hybrid strategies such as QUIK and SpQR, $1\%$ of the weights are maintained in $16$-bit. The underlined value shows the best quantization performance in each dataset, and the best averaged WER across all datasets is highlighted.}
\label{tab:whisper_tiny}
\begin{adjustbox}{width=\textwidth}
  \begin{tabular}{c c c c c c c c c c c}
  \toprule
  \multirow{2}{*}{Bits}  &  \multirow{2}{*}{Method} & \multicolumn{9}{c}{\textbf{Whisper Tiny (WER \%)} $\mathbf{\downarrow}$} \\
  \cmidrule(r){3-11}
  &  & AMI & Earnings-22 & GigaSpeech & Libri clean & Libri other & SPGISpeech & TED-Lium & Voxpopuli & \textbf{Avg.}\\
  \midrule
  Float &  Baseline &  24.66&	18.74&	14.12&	5.65&	15.44&	5.87&	5.98&	11.94&	12.80\\
  \midrule
  \multirow{8}{*}{w4-a8}  &	SmoothQuant & 47.25&	58.58&	30.94&	14.11&	37.55&	20.96&	20.08&	24.77&		31.78\\
  & AWQ & 								   38.77&	37.89&	20.67&	10.72&	20.82&	12.19&	12.56&	18.51&	21.52\\
  & OmniQuant & 						     37.14&	35.95&	21.66&	8.85&	20.95&	9.29&	12.35&	16.26&		20.31\\
  & GPTQ & 	      							     31.98&	33.91&	19.42&	9.06&	19.71&	10.18&	8.01&	16.30&		18.57\\
  & RTN &  								     38.51&	38.67&	21.11&	8.71&	21.07&	12.26&	13.62&	16.47&		21.30\\
  & TesseraQ & 							     30.51&	25.57&	17.93&	8.35&	20.16&	9.69&	9.59&	14.62&		17.05\\
  & QUIK & 								     \uline{28.17}&	\uline{24.21}&	\uline{16.05}&	\uline{6.50}&	\uline{16.15}&	\uline{7.40}&	\uline{6.35}&	\uline{12.69}&		\textbf{14.69}\\
  & SpQR & 								     29.63&	24.93&	16.71&	6.70&	17.99&	7.71&	8.56&	13.79&		15.75\\
  \midrule
  \multirow{8}{*}{w4-a16}  & SmoothQuant & 37.70&	42.94&	22.35&	8.75&	20.31&	12.38&	12.80&	17.99&		21.90\\
  & AWQ &   									31.42&	28.51&	18.75&	9.74&	19.21&	9.80&	9.88&	15.29&	17.83\\
  & OmniQuant &  							33.59&	26.06&	19.88&	7.46&	17.70&	8.36&	8.26&	13.49&		16.85\\
  & GPTQ & 									31.65&	27.94&	16.92&	7.07&	18.26&	8.20&	8.00&	13.60&		16.46\\
  & RTN & 									30.84&	33.28&	17.59&	7.69&	19.78&	9.79&	8.70&	14.54&		17.78\\
  & TesseraQ &  								30.59&	31.57&	17.80&	7.68&	19.82&	9.80&	8.86&	14.55&		17.58\\ %
  & QUIK &  									28.21&	24.56&	15.95&	6.35&	15.98&	7.37&	\uline{6.36}&	13.05&		14.73\\
  & SpQR &  								      \uline{26.54}&	\uline{24.47}&	\uline{15.62}&\uline{6.22}&	\uline{15.14}&	\uline{6.92}&	6.89&	\uline{12.11}&		\textbf{14.24}\\
  \midrule
  \multirow{8}{*}{w8-a8}  & SmoothQuant &   25.72&	23.95&	15.59&	6.09&	14.70&	6.74&	6.26&	12.36&		13.93\\
  & AWQ &  								25.37&	22.04&	15.03&	6.54&	15.74&	6.71&	7.75&	13.13&	14.04	\\
  & OmniQuant & 						      26.38&	21.27&	16.00&	6.12&	15.07&	6.69&	6.58&	12.75&		13.86\\
  & GPTQ &   							      25.80&	\uline{19.51}&	15.13&	6.02&	15.78&	6.70&	7.86&	12.44&		13.65\\
  & RTN &   								     24.82&	19.60&	14.89&	6.06&	16.04&	6.54&	6.27&	12.93&		13.39\\
  & TesseraQ &  								25.47&	19.95&	14.33&	6.25&	15.03&	6.26&	6.21&	\uline{11.75}&		13.16\\
  & QUIK & 									\uline{24.36}&	21.43&	\uline{13.94}&\uline{5.72}&	\uline{14.17}&	\uline{5.99}&	\uline{6.09}&	11.79&		\textbf{12.94}\\
  & SpQR & 									26.21&	22.01&	14.47&	5.97&	15.94&	6.75&	7.93&	13.03&		14.04\\
  \midrule
  \multirow{8}{*}{w8-a16}  & SmoothQuant & 24.36&	19.35&	13.97&	5.69&	14.64&	5.87&	5.99&	12.18&		12.76\\
  & AWQ &  								24.36&	19.66&	14.00&	5.68&	14.56&	\uline{5.83}&	5.97&	12.76&	12.85	\\
  & OmniQuant & 						      25.99&	19.17&	14.52&	5.68&	15.41&	5.90&	\uline{5.93}&	12.14&		13.09\\
  & GPTQ &   							      24.97&	20.19&	14.11&	5.67&	\uline{14.54}&	5.87&	5.96&	12.00&		12.91\\
  & RTN &   								      24.58&	18.65&	13.95&	5.67&	15.40&	5.90&	5.96&	12.16&		12.78\\
  & TesseraQ &  								24.68&	19.49&	\uline{13.84}&	5.70&	14.55&	5.91&	5.97&	\uline{11.33}&		\textbf{12.68}\\
  & QUIK & 									\uline{23.44}&	19.42&	14.00&	5.69&	15.40&	5.89&	5.99&	11.95&		12.72\\
  & SpQR & 									24.59&	\uline{18.93}&	13.89&	\uline{5.66}&	15.02&	5.88&	5.94&	11.94&		12.73\\
  \bottomrule
  \end{tabular}
\end{adjustbox}
\end{small}
\end{table}

\clearpage
\subsection{Whisper Base}

\begin{table}[h]
\centering
\begin{small}
\captionsetup{font=footnotesize}
\caption{Overall performance of quantized Whisper Base model across $7$ datasets under $8$ quantization algorithms. To align with typical hardware support, per-tensor symmetric quantization is used for activations, and per-group symmetric quantization is used for weights, unless otherwise specified. The group size for weights is $64$ for Whisper models. The underlined value shows the best quantization performance in each dataset, and the best averaged WER across all datasets is highlighted.}
\label{tab:whisper_base}
\begin{adjustbox}{width=\textwidth}
  \begin{tabular}{c c c c c c c c c c c}
  \toprule
  \multirow{2}{*}{Bits}  &  \multirow{2}{*}{Method} & \multicolumn{9}{c}{\textbf{Whisper Base (WER \%)} $\mathbf{\downarrow}$} \\
  \cmidrule(r){3-11}
  &  & AMI & Earnings-22 & GigaSpeech & Libri. clean & Libri. other & SPGISpeech & TED-Lium & Voxpopuli & \textbf{Avg.}\\
  \midrule
  Float &  Baseline &  21.13 & 15.09 & 12.87 & 4.28 &	10.36 & 4.27 & 4.85 & 9.75 & 10.32 \\
  \midrule
  \multirow{8}{*}{w4-a8}  &	SmoothQuant & 30.63&	22.50&	17.72&	12.31&	16.85&	8.16&	12.71&	13.63&		16.81\\
  & AWQ & 								     22.77&	18.95&	13.35&	5.16&	13.13&	5.01&	5.35&	11.23&	11.87\\
  & OmniQuant & 						     23.12&	17.53&	13.70&	4.68&	12.03&	5.23&	5.65&	12.23&		11.77\\
  & GPTQ & 	      							     22.69&	16.26&	13.09&	4.99&	12.17&	4.78&	5.60&	11.83&		11.43\\
  & RTN &  								     22.53&	18.05&	13.39&	4.59&	\uline{11.40}&	5.06&	5.51&	10.99&		11.44\\
  & TesseraQ & 								22.03&	16.79&	13.46&	5.20&	12.35&	5.06&	5.46&	10.89&		11.40\\
  & QUIK & 								     \uline{21.38} &	16.23&	13.52&	4.51&	12.72&	4.68&	5.33&	\uline{10.12}&		\textbf{11.06}\\
  & SpQR & 								    22.99&	\uline{15.66}&	\uline{13.01}& \uline{4.48}&	11.49&	\uline{4.66}&	\uline{5.02}&	11.84&		11.14\\
  \midrule
  \multirow{8}{*}{w4-a16}  & SmoothQuant & 27.13&	17.82&	13.76&	4.68&	13.30&	5.41&	6.54&	11.27&		12.49\\
  & AWQ &   									23.93&	\uline{15.86}&	13.76&	\uline{4.46}&	12.64&	4.88&	5.24&	\uline{10.48}&	11.41\\
  & OmniQuant &  							23.06&	18.00&	13.89&	4.64&	11.99&	5.05&	5.39&	10.71&		11.59\\
  & GPTQ & 									23.02&	16.22&	13.46&	4.55&	\uline{11.24}&	4.74&	5.43&	10.61&		11.16\\
  & RTN & 									23.77&	16.64&	13.37&	4.52&	11.63&	4.88&	5.27&	10.83&		11.36\\
  & TesseraQ &  								23.35&	17.53&	13.29&	4.52&	11.66&	4.92&	5.31&	10.78&		11.42\\ %
  & QUIK &  									\uline{21.84}&	16.84&	13.80&	4.62&	12.34&	4.74&	5.26&	10.99&		11.30\\
  & SpQR &  								      22.25&	16.49&	\uline{12.47}&	\uline{4.46}&	11.34&	\uline{4.59}&	\uline{4.93}&	11.00&		\textbf{10.94}\\
  \midrule
  \multirow{8}{*}{w8-a8}  & SmoothQuant &   21.56&	16.68&	12.80&	4.57&	11.03&	4.58&	5.02&	10.29&		10.82\\
  & AWQ &  								21.69&	\uline{15.02}&	12.56&	4.22&	10.68&	4.39&	4.89&	9.74&	10.40 \\ %
  & OmniQuant & 						      22.52&	16.61&	13.02&	4.22&	10.98&	4.62&	5.06&	13.06&		11.26\\
  & GPTQ &   							      21.02&	15.22&	12.84&	4.32&	11.47&	4.38&	5.01&	9.89&		10.52\\
  & RTN &   								     21.92&	16.05&	12.31&	4.27&	10.39&	4.37&	5.00&	9.79&		10.51\\
  & TesseraQ &  								\uline{20.68}&	16.05&	12.43&	\uline{4.18}&	11.51&	4.30&	\uline{4.87}&	10.71&		10.59\\
  & QUIK & 									21.07&	\uline{15.02}&\uline{12.12}&	4.22&	\uline{10.13}&	\uline{4.24}&	4.94&	\uline{9.71}&		\textbf{10.18}\\
  & SpQR & 									21.35&	16.37&	12.50&	4.25&	10.66&	4.44&	4.93&	9.84&		10.54\\
  \midrule
  \multirow{8}{*}{w8-a16}  & SmoothQuant & 21.50&	15.05&	12.73&	4.28&	10.45&	4.28&	4.90&	10.74&		10.49\\
  & AWQ &  									20.54&	15.14&	12.54&	4.25&	10.27&	4.27&	4.89&	9.78&	10.21\\
  & OmniQuant & 						      21.37&	15.38&	12.82&	4.26&	10.45&	4.33&	\uline{4.87}&	9.77&		10.41\\
  & GPTQ &   							      21.58&	15.04&	12.74&	4.29&	10.63&	4.26&	4.89&	9.79&		10.40\\
  & RTN &   								      21.15&	15.05&	12.59&	4.25&	10.46&	4.28&	4.89&	\uline{9.71}&		10.30\\
  & TesseraQ &  								\uline{20.53}&	\uline{15.03}&	\uline{12.41}&	4.26&	10.19&	4.27&	4.92&	9.78&		\textbf{10.17}\\
  & QUIK & 									21.13&	15.08&	12.69&	4.28&	\uline{10.16}&\uline{4.23}&	4.91&	9.78&		10.28\\
  & SpQR & 									20.76&	15.10&	12.63&	\uline{4.24}&	10.38&	4.26&	4.90&	9.75&		10.25 \\ 
  \bottomrule
  \end{tabular}
\end{adjustbox}
\end{small}
\end{table}

\clearpage 

\subsection{Whisper Small}
\begin{table}[h]
\centering
\captionsetup{font=footnotesize}
\begin{small}
\caption{Overall performance of quantized Whisper Small model across $7$ datasets under $8$ quantization algorithms. To align with typical hardware support, per-tensor symmetric quantization is used for activations, and per-group symmetric quantization is used for weights, unless otherwise specified. The group size for weights is $64$ for Whisper models. The underlined value shows the best quantization performance in each dataset, and the best averaged WER across all datasets is highlighted.}
\label{tab:whisper_small}
\begin{adjustbox}{width=\textwidth}
  \begin{tabular}{c c c c c c c c c c c}
  \toprule
  \multirow{2}{*}{Bits}  &  \multirow{2}{*}{Method} & \multicolumn{9}{c}{\textbf{Whisper Small (WER \%)} $\mathbf{\downarrow}$} \\
  \cmidrule(r){3-11}
  &  & AMI & Earnings-22 & GigaSpeech & Libri clean & Libri other & SPGISpeech & TED-Lium & Voxpopuli & \textbf{Avg.}\\
  \midrule
  Float &  Baseline &  17.95&	12.99&	11.36&	3.03&	7.27&	3.58&	4.06&	8.51&		8.59\\
  \midrule
  \multirow{8}{*}{w4-a8}  &	SmoothQuant & 18.29&	14.15&	12.46&	4.08&	9.41&	4.15&	4.34&	10.57&		9.68\\
  & AWQ & 								     18.51&	13.14&	11.10&	3.18&	7.92&	3.82&	4.38&	8.52&	8.82\\
  & OmniQuant & 						     \uline{17.68}&	13.19&	11.33&	3.06&	7.40&	3.70&	\uline{4.09}&	8.69&		8.64\\
  & GPTQ & 	      							     18.09&	13.11&	11.44&	3.07&	7.86&	3.67&	4.20&	\uline{8.47}&		8.74\\
  & RTN &  								     18.46&	13.28&	11.16&	3.19&	7.34&	3.88&	4.26&	8.65&		8.78\\
  & TesseraQ & 								17.96&	13.40&	\uline{11.00}&	3.10&	7.78&	3.84&	4.15&	8.61&		8.73\\
  & QUIK & 								     17.81&	14.21&	11.28&	3.21&	7.47&	3.91&	4.36&	8.60&		8.86 \\
  & SpQR & 								     18.30&	\uline{12.95}&	11.29&	\uline{3.00}&	\uline{7.10}&	\uline{3.49}&	4.13&	8.56&		\textbf{8.60}\\
  \midrule
  \multirow{8}{*}{w4-a16}  & SmoothQuant & 18.03&	13.52&	12.11&	3.33&	7.81&	4.17&	4.36&	8.64&		9.00\\
  & AWQ &   									17.87&	14.14&	11.48&	3.29&	7.52&	4.04&	4.36&	8.69&	8.92\\
  & OmniQuant &  							18.59&	13.85&	11.54&	3.21&	7.28&	\uline{3.73}&	4.20&	8.98&		8.92\\
  & GPTQ & 									18.43&	13.36&	11.23&	3.10&	7.40&	3.83&	\uline{4.19}&	8.65&		8.77\\
  & RTN & 									17.86&	13.60&	11.45&	3.37&	7.44&	4.17&	4.31&	8.72&		8.86\\
  & TesseraQ &  								17.83&	13.63&	11.57&	3.40&	7.36&	4.16&	4.29&	8.71&		8.87\\ %
  & QUIK &  									\uline{17.77}&	14.31&	11.53&	3.26&	7.41&	3.93&	4.38&	\uline{8.59}&		8.90\\
  & SpQR &  								      17.96&	\uline{13.17}&	\uline{11.13}&	\uline{3.02}&	\uline{7.15}&	3.75&	4.30&	8.67&		\textbf{8.65}\\
  \midrule
  \multirow{8}{*}{w8-a8}  & SmoothQuant &   17.99&	13.05&	11.11&	3.08&	7.29&	3.56&	4.00&	8.33&		8.55\\
  & AWQ &  									17.67&	13.00&	11.44&	3.06&	7.53&	3.75&	4.22&	8.42&	8.64\\
  & OmniQuant & 						      \uline{17.37}&	12.99&	\uline{11.05}&	2.99&	\uline{7.10}&	\uline{3.48}&	4.10&	8.46&		\textbf{8.44}\\
  & GPTQ &   							      17.59&	\uline{12.92}&	11.13&	\uline{2.95}&	7.13&	3.49&	4.04&	8.32&		8.45\\
  & RTN &   								     18.05&	12.96&	11.18&	2.96&	7.23&	3.52&	4.14&	8.45&		8.56\\
  & TesseraQ &  								17.53&	12.97&	11.14&	2.99&	7.22&	\uline{3.48}&	\uline{3.95}&	8.53&		8.47\\
  & QUIK & 									18.51&	12.97&	11.30&	3.00&	7.16&	3.59&	4.02&	8.48&		8.63\\
  & SpQR & 									18.02&	12.96&	11.14&	2.97&	7.43&	3.54&	3.98&	\uline{8.30}&		8.54\\
  \midrule
  \multirow{8}{*}{w8-a16}  & SmoothQuant & 17.95&	13.03&	11.59&	3.06&	7.25&	3.59&	4.05&	8.48&		8.62\\
  & AWQ &  									\uline{17.72}&	\uline{12.78}&\uline{11.24}&	\uline{2.99}&	\uline{7.13}&	\uline{3.56}&	\uline{3.91}&	8.56&	\textbf{8.49}\\
  & OmniQuant & 						      17.93&	13.05&	11.32&	3.04&	7.17&	3.62&	4.05&	8.48&		8.58\\
  & GPTQ &   							      17.93&	13.00&	11.44&	3.08&	7.24&	3.63&	4.05&	8.51&		8.61\\
  & RTN &   								      17.95&	12.96&	11.36&	3.06&	7.55&	3.64&	4.10&	8.48&		8.64\\
  & TesseraQ &  								17.93&	13.04&	11.52&	3.04&	7.47&	3.60&	4.08&	\uline{8.47}&		8.64\\
  & QUIK & 									17.91&	12.98&	11.41&	3.06&	7.46&	3.63&	4.05&	8.49&		8.62\\
  & SpQR & 									17.91&	12.98&	11.36&	3.06&	7.25&	3.58&	4.08&	\uline{8.47}&		8.59\\
  \bottomrule
  \end{tabular}
\end{adjustbox}
\end{small}
\end{table}

\clearpage 

\subsection{Moonshine Tiny}
\begin{table}[h]
\centering
\captionsetup{font=footnotesize}
\begin{small}
\caption{Overall performance of quantized Moonshine Tiny model across $7$ datasets under $8$ quantization algorithms. To align with typical hardware support, per-tensor symmetric quantization is used for activations, and per-group symmetric quantization is used for weights, unless otherwise specified. The group size for weights is $72$ for Moonshine Tiny. The underlined value shows the best quantization performance in each dataset, and the best averaged WER across all datasets is highlighted.}
\label{tab:moonshine_tiny}
\begin{adjustbox}{width=\textwidth}
  \begin{tabular}{c c c c c c c c c c c}
  \toprule
  \multirow{2}{*}{Bits}  &  \multirow{2}{*}{Method} & \multicolumn{9}{c}{\textbf{Moonshine Tiny} (WER \%) $\mathbf{\downarrow}$} \\
  \cmidrule(r){3-11}
  &  & AMI & Earnings-22 & GigaSpeech & Libri clean & Libri other & SPGISpeech & TED-Lium & Voxpopuli & \textbf{Avg.}\\
  \midrule
  Float &  Baseline &  22.41&	22.00&	14.19&	4.60&	11.84&	7.43&	5.68&	13.58&		12.72\\
  \midrule
  \multirow{8}{*}{w4-a8}  &	SmoothQuant & 90.12&	108.42&	58.50&	45.68&	82.93&	88.83&	35.87&	73.57&		72.99\\
  & AWQ & 								    26.90&	25.76&	16.65&	5.73&	14.97&	9.77&	7.04&	15.73&	15.32\\
  & OmniQuant & 						     25.41&	24.97&	15.87&	5.41&	14.36&	9.53&	7.08&	15.27&		14.74\\
  & GPTQ & 	      							    \uline{24.65}&	\uline{24.24}&	15.64&	5.37&	\uline{13.47}&	8.94&	6.54&	14.32&		\textbf{14.15}\\
  & RTN &  								     40.34&	39.43&	24.78&	11.36&	30.67&	21.59&	11.77&	23.07&		25.38\\
  & TesseraQ & 							     30.36&	31.74&	19.07&	6.95&	17.57&	12.84&	8.64&	16.43&		17.95\\
  & QUIK & 								     25.18&	24.82&	\uline{15.31}&	5.54&	14.17&	\uline{8.46}&	\uline{6.30}&	14.80&		14.32\\
  & SpQR & 								     25.07&	24.42&	15.64&	\uline{5.28}&	13.70&	8.69&	6.47&	\uline{13.98}&		\textbf{14.15}\\
  \midrule
  \multirow{8}{*}{w4-a16}  & SmoothQuant & 88.74&	107.73&	53.25&	41.96&	77.65&	80.95&	34.97&	73.26&		69.81\\
  & AWQ &   									26.38&	26.13&	16.05&	5.70&	14.64&	9.38&	7.04&	15.63&	15.12\\
  & OmniQuant &  							24.68&	24.97&	15.45&	5.41&	13.90&	9.18&	7.04&	15.15&		14.47\\
  & GPTQ & 									\uline{23.79}&	\uline{23.27}&	15.03&	5.09&	\uline{13.01}&	8.22&	6.37&	14.12&		\textbf{13.61}\\
  & RTN & 									37.72&	38.64&	22.13&	9.37&	25.39&	18.12&	11.88&	22.55&		23.23\\
  & TesseraQ &  								26.30&	26.65&	16.11&	5.78&	14.66&	9.47&	7.45&	15.24&		15.21\\
  & QUIK &  									25.03&	24.64&	15.19&	6.00&	13.59&	8.35&	\uline{6.21}&	14.88&		14.24\\
  & SpQR &  								      23.91&	24.20&	\uline{14.99}&	\uline{5.07}&	13.12&	\uline{8.18}&	6.32&	\uline{14.66}&		13.81\\
  \midrule
  \multirow{8}{*}{w8-a8}  & SmoothQuant &   23.32&	22.43&	15.10&	5.10&	12.63&	8.18&	6.15&	14.45&		13.42\\
  & AWQ &  									23.25&	22.34&	14.71&	4.82&	12.32&	7.94&	5.78&	13.78&	13.12\\
  & OmniQuant & 						     23.15&	22.40&	14.48&	4.72&	12.18&	7.66&	5.97&	13.78&		13.04\\
  & GPTQ &   							     23.15&	22.64&	14.64&	4.77&	12.15&	7.79&	5.80&	13.81&		13.09\\
  & RTN &   								      23.11&	22.28&	14.74&	4.74&	12.19&	7.79&	5.77&	13.80&		13.05\\
  & TesseraQ &  								23.93&	23.55&	14.91&	4.82&	12.79&	8.05&	6.04&	13.93&		13.50\\
  & QUIK & 									\uline{22.63}&	22.34&	\uline{14.30}&	\uline{4.61}&	\uline{11.88}&	\uline{7.53}&	\uline{5.67}&	\uline{13.53}&		\textbf{12.81}\\
  & SpQR & 									23.17&	\uline{22.23}&	14.71&	4.75&	12.15&	7.77&	5.76&	13.75&		13.04\\
  \midrule
  \multirow{8}{*}{w8-a16}  & SmoothQuant & 22.47&	21.99&	14.19&	4.60&	12.00&	7.46&	5.68&	13.81&		12.78\\
  & AWQ &  									22.45&	22.12&	14.19&	4.58&	11.94&	7.46&	5.67&	13.80&	12.78\\
  & OmniQuant & 						      \uline{22.34}&	\uline{21.96}&	14.21&	4.61&	11.90&	7.44&	5.63&	13.76&		\textbf{12.73}\\
  & GPTQ &   							      22.47&	22.11&	\uline{14.16}&	4.61&	\uline{11.87}&	7.45&	5.69&	13.80&		12.77\\
  & RTN &   								      22.40&	22.08&	14.19&	4.62&	11.95&	7.44&	\uline{5.61}&	13.79&		12.76\\
  & TesseraQ &  								22.39&	22.10&	14.17&	4.61&	11.88&	7.43&	5.64&	13.72&		12.74\\
  & QUIK & 									22.42&	22.12&	14.20&	\uline{4.57}&	11.88&	\uline{7.42}&	5.65&	13.65&		12.74\\
  & SpQR & 									22.48&	22.18&	14.20&	\uline{4.57}&	11.89&	7.44&	5.69&	\uline{13.63}&		12.76\\
  \bottomrule
  \end{tabular}
\end{adjustbox}
\end{small}
\end{table}

\clearpage 

\subsection{Moonshine Base}
\begin{table}[h]
\centering
\captionsetup{font=footnotesize}
\begin{small}
\caption{Overall performance of quantized Moonshine Base model across $7$ datasets under $8$ quantization algorithms. To align with typical hardware support, per-tensor symmetric quantization is used for activations, and per-group symmetric quantization is used for weights, unless otherwise specified. The group size for weights is $52$ for Moonshine Base. The underlined value shows the best quantization performance in each dataset, and the best averaged WER across all datasets is highlighted.}
\label{tab:moonshine_base}
\begin{adjustbox}{width=\textwidth}
  \begin{tabular}{c c c c c c c c c c c}
  \toprule
  \multirow{2}{*}{Bits}  &  \multirow{2}{*}{Method} & \multicolumn{9}{c}{\textbf{Moonshine Base} (WER \%) $\mathbf{\downarrow}$} \\
  \cmidrule(r){3-11}
  &  & AMI & Earnings-22 & GigaSpeech & Libri clean & Libri other & SPGISpeech & TED-Lium & Voxpopuli & \textbf{Avg.}\\
  \midrule
  Float &  Baseline &  17.07 &	17.69 &	12.11 &	3.26 &	8.28 &	5.46 &	5.24 &	10.79 &		9.99\\
  \midrule
  \multirow{8}{*}{w4-a8}  &	SmoothQuant & 84.92&	41.95&	102.77&	106.44&	106.00&	105.48&	29.45&	40.76&		77.22\\
  & AWQ & 								   18.47&	19.72&	12.84&	3.77&	9.46&	6.05&	5.83&	11.74&	10.98\\
  & OmniQuant & 						     18.32&	19.36&	12.61&	3.60&	9.12&	5.74&	5.46&	11.68&		10.74\\
  & GPTQ & 	      							     17.70&	18.31&	12.43&	3.50&	9.01&	5.76&	\uline{5.21}&	11.12&		10.38\\
  & RTN &  								     22.64&	25.07&	14.29&	4.50&	12.20&	8.21&	6.76&	14.21&		13.48\\
  & TesseraQ & 							     21.41&	22.05&	13.49&	3.95&	10.67&	6.95&	6.04&	12.04&		12.08\\
  & QUIK & 								     17.81&	18.80&	\uline{12.39}&\uline{3.40}&	\uline{8.83}&	\uline{5.54}&	5.22&	10.99&		10.37\\
  & SpQR & 								     \uline{17.58}&	\uline{17.93}&	12.46&	3.43&	9.13&	5.83&	\uline{5.21}&	\uline{10.87}&		\textbf{10.31}\\
  \midrule
  \multirow{8}{*}{w4-a16}  & SmoothQuant & 22.43&	22.94&	14.22&	5.55&	15.77&	13.78&	6.33&	12.57&		14.20\\
  & AWQ &   									17.94&	19.60&	12.59&	3.46&	8.98&	\uline{5.43}&	5.63&	10.96&	10.57\\
  & OmniQuant &  							17.88&	19.16&	12.50&	3.51&	8.87&	5.66&	5.48&	11.25&		10.54\\
  & GPTQ & 									17.39&	18.31&	\uline{12.31}&	3.41&	\uline{8.74}&	5.58&	5.29&	11.25&		10.28\\
  & RTN & 									21.80&	24.12&	13.81&	4.53&	11.55&	7.50&	6.71&	13.50&		12.94\\
  & TesseraQ &  								18.63&	19.50&	12.61&	3.62&	9.30&	5.73&	5.19&	11.70&		10.78\\
  & QUIK &  									17.73&	18.77&	12.36&	3.49&	8.97&	5.55&	5.27&	11.04&		10.40\\
  & SpQR &  								      \uline{17.20}&	\uline{17.82}&	12.39&	\uline{3.30}&	8.76&	5.51&	\uline{5.24}&	\uline{10.89}&		\textbf{10.14}\\
  \midrule
  \multirow{8}{*}{w8-a8}  & SmoothQuant &   32.81&	19.62&	45.83&	66.87&	60.45&	65.74&	6.58&	13.07&		38.87\\
  & AWQ &  									17.30&	17.92&	12.14&	3.46&	8.55&	5.66&	5.13&	11.10&	10.16\\
  & OmniQuant & 						      17.51&	18.42&	12.14&	3.33&	8.55&	\uline{5.44}&	\uline{4.89}&	10.88&		10.14\\
  & GPTQ &   							      17.24&	17.95&	12.26&	3.37&	8.58&	5.66&	5.22&	10.95&		10.15\\
  & RTN &   								      17.22&	17.81&	12.20&	3.36&	8.62&	5.70&	4.99&	\uline{10.79}&		10.09\\
  & TesseraQ &  								17.84&	18.18&	12.30&	3.54&	8.76&	5.85&	5.00&	10.87&		10.29\\
  & QUIK & 									\uline{17.19}&	\uline{17.77}&	\uline{12.13}&\uline{3.30}&	\uline{8.34}&	5.45&	5.13&	11.01&		\textbf{10.04}\\
  & SpQR & 									17.26&	17.82&	12.24&	3.36&	8.52&	5.73&	5.14&	10.88&		10.12\\
  \midrule
  \multirow{8}{*}{w8-a16}  & SmoothQuant & \uline{16.98}&	17.74&	12.12&	3.30&	\uline{8.25}&	5.46&	5.23&	\uline{10.78}&		\textbf{9.98}\\
  & AWQ &  									17.08&	17.97&	\uline{12.09}&	\uline{3.25}&	8.30&	5.48&	5.25&	11.03&	10.06\\
  & OmniQuant & 						      17.17&	17.73&	12.10&	3.29&	8.31&	\uline{5.42}&	5.34&	11.05&		10.05\\
  & GPTQ &   							      17.10&	17.71&	12.10&	3.27&	8.29&	5.47&	\uline{5.17}&	10.83&		9.99\\
  & RTN &   								      17.11&	\uline{17.62}&	12.10&	3.29&	8.27&	5.50&	5.28&	11.07&		10.03\\
  & TesseraQ &  								17.06&	17.84&	12.12&	3.27&	8.33&	5.47&	5.27&	11.07&		10.05\\
  & QUIK & 									17.12&	17.69&	12.10&	3.28&	8.29&	5.47&	5.31&	11.05&		10.04\\
  & SpQR & 									17.09&	17.75&	12.10&	3.27&	8.27&	5.48&	\uline{5.17}&	11.08&		10.03\\
  \bottomrule
  \end{tabular}
\end{adjustbox}
\end{small}
\end{table}

\subsection{Runtime Cost of Quantization Algorithms}
\label{appendix:runtime}
\begin{table}[h]
\centering
\begin{small}
\captionsetup{font=footnotesize}
\caption{Runtime comparison of diverse quantization algorithms. Numbers are reported in seconds, and are based on the quantization process for Whisper Base and Moonshine Base models with w4-a16 configuration, running on single Nivida GPU - RTX 4090 with memory size of $24$G. $256$ samples are used for calibration with batch size of $1$. Among the $8$ quantization processes, AWQ consumes the most time for calibration.}
\label{tab:runtime}
\scalebox{0.9}{  
\begin{tabular}{c c c c c c c c c}
  \toprule
  Runtime (sec.) & SmoothQuant & AWQ & OmniQuant & GPTQ & RTN & TesseraQ & QUIK & SpQR \\
   \midrule
  Whisper Base & 30.30 & 33409.44 & 773.13 & 85.06 & 0.78 & 331.12 & 21.79 & 69.83 \\
   \midrule
  Moonshine Base & 32.31 & 44360.54 & 1035.37 &99.36 & 0.76 & 978.33 & 25.28 & 90.36 \\
\bottomrule
  \end{tabular}}
\end{small}
\end{table}

\clearpage
\subsection{Quantization Granularity Exeperiments}
\label{appendix:granularity}

\begin{table}[h]
\centering
\begin{small}
\captionsetup{font=scriptsize} 
\caption{In addition to Table \ref{tab:granularity}, per-dataset WER is reported for granularity experiments. For per-group weight quantization, the coarse group sizes for Whisper, Moonshine Tiny and Moonshine Base are $128$, $144$ and $104$, and the fine group sizes are $64$, $72$ and $52$, respectively. Activations are using $16$-bit per-tensor or per-token quantization, and weights are using $4$-bit per-channel or per-group quantization. The best averaged WER across all datasets is highlighted.}
\label{tab:granularity_perdataset}
\scalebox{0.9}{
\begin{adjustbox}{width=\textwidth}
  \begin{tabular}{c c c c c c c c c c c}
  \toprule
  Granularity w/a & Method & AMI & Earnings-22 & GigaSpeech & Libri clean & Libri other & SPGISpeech & TED-Lium & Voxpopuli & \textbf{Avg.}\\
  \midrule
 \multicolumn{11}{c}{\textbf{Whisper Tiny w4-a16 (WER \%)} $\mathbf{\downarrow}$} \\
  \midrule
  & Float &  24.66&	18.74&	14.12&	5.65&	15.44&	5.87&	5.98&	11.94&	12.80\\
  \midrule
  \multirow{3}{*}{per-channel/per-tensor} & AWQ & 65.96&	94.10&	55.17&	41.81&	96.34&	58.22&	37.31&	65.80&	64.34\\
  & GPTQ & 47.63&	53.13&	26.78&	13.94&	33.99&	15.78&	11.54&	21.52&		\textbf{28.04}\\
  & TesseraQ & 55.04&	81.12&	36.94&	26.08&	46.40&	35.19&	22.32&	45.54&		43.58\\
\midrule
   \multirow{3}{*}{per-group, coarse/per-tensor} & AWQ & 41.04&	46.36&	23.66&	8.61&	22.89&	13.95&	10.75&	18.84&	23.26\\
  & GPTQ & 	31.52&	26.25&	18.23&	7.74&	19.19&	9.04&	8.25&	14.48&		\textbf{16.84}\\
  & TesseraQ & 35.53&	40.02&	19.78&	8.50&	20.93&	12.42&	8.93&	17.94&		20.51\\  
  \midrule
   \multirow{3}{*}{per-group, fine/per-tensor} & AWQ &  31.42&	28.51&	18.75&	9.74&	19.21&	9.80&	9.88&	15.29&	17.83\\
  & GPTQ & 31.65&	27.94&	16.92&	7.07&	18.26&	8.20&	8.00&	13.60&		\textbf{16.46}\\
  & TesseraQ & 30.59&	31.57&	17.80&	7.68&	19.82&	9.80&	8.86&	14.55&		17.58 \\  
\midrule
   \multirow{3}{*}{per-group, fine/per-token} & AWQ & 31.51&	29.60&	18.61&	9.74&	19.26&	9.79&	8.77&	15.26&	17.82\\
  & GPTQ & 33.46&	28.91&	17.75&	7.56&	16.82&	8.29&	8.08&	12.92&		\textbf{16.72}\\
  & TesseraQ & 30.84&	33.28&	17.59&	7.69&	19.78&	9.79&	8.70&	14.54&		17.78\\
\midrule
 \multicolumn{11}{c}{\textbf{Whisper Base w4-a16 (WER \%)} $\mathbf{\downarrow}$} \\
  \midrule
& Float &  21.13&	15.09&	12.87&	4.28&	10.36&	4.27&	4.85&	9.75&	10.32\\
  \midrule
  \multirow{3}{*}{per-channel/per-tensor} & AWQ & 65.54&	93.77&	63.01&	91.68&	128.17&	57.10&	82.33&	103.53&	85.64\\
  & GPTQ & 31.12&	26.66&	16.70&	5.28&	15.14&	6.70&	7.88&	13.08&		\textbf{15.32}\\
  & TesseraQ & 32.31&	33.22&	17.03&	8.93&	20.10&	8.95&	11.25&	12.65&		18.06\\
\midrule
   \multirow{3}{*}{per-group, coarse/per-tensor} & AWQ & 25.20&	18.97&	14.81&	4.71&	13.30&	5.38&	5.61&	13.13&	12.64\\
  & GPTQ & 	23.39&	16.89&	13.67&	4.65&	11.57&	5.08&	5.54&	12.05&		\textbf{11.60}\\
  & TesseraQ & 23.28&	18.09&	13.32&	5.11&	12.45&	5.26&	5.54&	10.91&		11.75\\  
  \midrule
   \multirow{3}{*}{per-group, fine/per-tensor} & AWQ & 23.93&	15.86&	13.76&	4.46&	12.64&	4.88&	5.24&	10.48&	11.41\\
  & GPTQ & 23.02&	16.22&	13.46&	4.55&	11.24&	4.74&	5.43&	10.61&		\textbf{11.16}\\
  & TesseraQ &  23.35&	17.53&	13.29&	4.52&	11.66&	4.92&	5.31&	10.78&		11.42\\  %
\midrule
   \multirow{3}{*}{per-group, fine/per-token} & AWQ & 23.68&	15.87&	14.01&	4.48&	12.24&	4.94&	5.25&	10.56&	11.38\\
  & GPTQ & 25.49&	16.68&	13.56&	4.64&	12.03&	4.97&	5.44&	12.25&		11.88\\
  & TesseraQ & 23.77&	16.64&	13.37&	4.52&	11.63&	4.88&	5.27&	10.83&		\textbf{11.36} \\
\midrule
 \multicolumn{11}{c}{\textbf{Whisper Small w4-a16 (WER \%)} $\mathbf{\downarrow}$} \\
  \midrule
& Float & 17.95&	12.99&	11.36&	3.03&	7.27&	3.58&	4.06&	8.51&	8.59 \\
  \midrule
  \multirow{3}{*}{per-channel/per-tensor} & AWQ & 20.59&	14.46&	12.23&	4.01&	9.52&	4.70&	4.45&	9.59&	9.94\\
  & GPTQ & 18.32&	14.61&	12.14&	3.36&	8.24&	4.24&	4.23&	9.01&		\textbf{9.27}\\
  & TesseraQ & 19.18&	13.59&	11.77&	3.67&	9.24&	4.36&	4.47&	9.13&		9.42\\
\midrule
   \multirow{3}{*}{per-group, coarse/per-tensor} & AWQ & 18.05&	13.39&	11.38&	3.16&	7.65&	4.21&	4.40&	8.68&	8.86\\
  & GPTQ & 	17.92&	13.36&	11.68&	3.17&	7.54&	3.80&	4.21&	8.78&		\textbf{8.81}\\
  & TesseraQ & 17.88&	13.87&	11.56&	3.49&	7.53&	4.23&	4.40&	8.74&		8.96\\  
  \midrule
   \multirow{3}{*}{per-group, fine/per-tensor} & AWQ & 17.87&	14.14&	11.48&	3.29&	7.52&	4.04&	4.36&	8.69&	8.92\\
  & GPTQ & 18.43&	13.36&	11.23&	3.10&	7.40&	3.83&	4.19&	8.65&		\textbf{8.77}\\
  & TesseraQ & 17.83&	13.63&	11.57&	3.40&	7.36&	4.16&	4.29&	8.71&		8.87 \\  %
\midrule
   \multirow{3}{*}{per-group, fine/per-token} & AWQ & 17.79&	14.16&	11.47&	3.35&	7.62&	4.22&	4.37&	8.67&	8.96\\
  & GPTQ &18.99&	13.49&	11.51&	3.56&	7.36&	3.80&	4.09&	8.79&		8.95 \\
  & TesseraQ & 17.86&	13.60&	11.45&	3.37&	7.44&	4.17&	4.31&	8.72&		\textbf{8.86}\\
\midrule
 \multicolumn{11}{c}{\textbf{Moonshine Tiny w4-a16 (WER \%)} $\mathbf{\downarrow}$} \\
  \midrule
& Float &  22.41&	22.00&	14.19&	4.60&	11.84&	7.43&	5.68&	13.58&	12.72\\
  \midrule
  \multirow{3}{*}{per-channel/per-tensor} & AWQ & 27.25&	27.08&	16.53&	6.02&	15.37&	9.99&	7.60&	15.51&	15.67\\
  & GPTQ & 25.70&	25.22&	15.63&	5.95&	14.16&	8.60&	6.75&	14.80&		\textbf{14.60}\\
  & TesseraQ & 27.79&	27.94&	17.10&	6.07&	17.04&	9.96&	7.95&	16.27&		16.26\\
\midrule
   \multirow{3}{*}{per-group, coarse/per-tensor} & AWQ & 25.95&	26.11&	16.14&	5.86&	15.12&	9.37&	7.22&	15.62&	15.17\\
  & GPTQ & 	24.39&	23.99&	15.31&	5.14&	13.41&	8.27&	6.53&	14.24&		\textbf{13.91}\\
  & TesseraQ & 26.98&	27.05&	16.68&	5.88&	15.34&	9.96&	7.41&	15.92&		15.65\\  
  \midrule
   \multirow{3}{*}{per-group, fine/per-tensor} & AWQ & 26.38&	26.13&	16.05&	5.70&	14.64&	9.38&	7.04&	15.63&	15.12\\
  & GPTQ & 23.79&	23.27&	15.03&	5.09&	13.01&	8.22&	6.37&	14.12&		\textbf{13.61}\\
  & TesseraQ & 26.30&	26.65&	16.11&	5.78&	14.66&	9.47&	7.45&	15.24&		15.21\\ 
\midrule
   \multirow{3}{*}{per-group, fine/per-token} & AWQ & 26.10&	25.85&	15.84&	5.48&	14.51&	9.22&	7.00&	15.57&	14.95\\
  & GPTQ & 23.86&	23.35&	14.94&	5.07&	13.00&	8.36&	6.35&	14.42&		\textbf{13.67}\\
  & TesseraQ & 26.42&	26.71&	16.17&	5.76&	14.68&	9.43&	7.49&	15.26&		15.24\\
\midrule
 \multicolumn{11}{c}{\textbf{Moonshine Base w4-a16 (WER \%)} $\mathbf{\downarrow}$} \\
  \midrule
  & Float &  17.07&	17.69&	12.11&	3.26&	8.28&	5.46&	5.24&	10.79&	9.99\\
  \midrule
  \multirow{3}{*}{per-channel/per-tensor} & AWQ & 19.03&	20.65&	12.97&	3.69&	9.56&	5.93&	5.40&	12.06&	11.16\\
  & GPTQ & 17.93&	19.55&	12.61&	3.52&	9.02&	6.05&	5.79&	11.44&		\textbf{10.74}\\
  & TesseraQ & 19.65&	20.59&	13.11&	3.86&	9.90&	6.27&	5.66&	12.41&		11.43\\
\midrule
   \multirow{3}{*}{per-group, coarse/per-tensor} & AWQ & 18.18&	20.16&	12.60&	3.59&	9.19&	5.86&	5.21&	11.53&	10.79\\
  & GPTQ & 17.93&	18.78&	12.60&	3.43&	8.89&	5.70&	5.31&	11.38&		\textbf{10.50}	\\
  & TesseraQ & 18.83&	20.77&	12.84&	3.58&	9.75&	6.22&	5.35&	12.25&		11.20\\  
  \midrule
   \multirow{3}{*}{per-group, fine/per-tensor} & AWQ & 17.94&	19.60&	12.59&	3.46&	8.98&	5.43&	5.63&	10.96&	10.57\\
  & GPTQ & 17.39&	18.31&	12.31&	3.41&	8.74&	5.58&	5.29&	11.25&		\textbf{10.28}\\
  & TesseraQ & 18.63&	19.50&	12.61&	3.62&	9.30&	5.73&	5.19&	11.70&		10.78\\  %
\midrule
   \multirow{3}{*}{per-group, fine/per-token} & AWQ & 18.07&	19.39&	12.62&	3.46&	8.89&	5.54&	5.62&	11.57&	10.64\\
  & GPTQ & 17.60&	18.42&	12.19&	3.45&	8.67&	5.47&	5.45&	11.14&		\textbf{10.30}\\
  & TesseraQ & 18.54&	19.34&	12.63&	3.65&	9.36&	5.73&	5.18&	11.69&		10.77\\
\bottomrule
 \end{tabular}
\end{adjustbox}}
\end{small}
\end{table}

\clearpage
\subsection{Weight Clipping Exeperiments}
\label{appendix:weightclipping}

\begin{table}[h]
\centering
\begin{small}
\captionsetup{font=footnotesize}
\caption{In addition to Table \ref{tab:weigt_clipping}, per-dataset WER is reported for weight clipping experiments. Per-tensor quantization is used for activations, and per-group quantization is used for weights, with a group size of $64$. Both weights and activations are using symmetric quantization. The best averaged WER across all datasets is highlighted.}
\label{tab:weightclipping_perdataset}
\begin{adjustbox}{width=\textwidth}
  \begin{tabular}{c c c c c c c c c c c}
  \toprule
  Bits & Method & AMI & Earnings-22 & GigaSpeech & Libri clean & Libri other & SPGISpeech & TED-Lium & Voxpopuli & \textbf{Avg.}\\
  \midrule
 \multicolumn{11}{c}{\textbf{Whisper Tiny (WER \%)} $\mathbf{\downarrow}$} \\
  \midrule
  & Float &  24.66&	18.74&	14.12&	5.65&	15.44&	5.87&	5.98&	11.94&	12.80\\
  \midrule
\multirow{2}{*}{w4-a8} & w/ clip & 107.70&	83.45&	48.37&	11.59&	44.29&	19.48&	10.89&	23.61&	43.67 \\
& w/o clip & 38.77&	37.89&	20.67&	10.72&	20.82&	12.19&	12.56&	18.51&	\textbf{21.52} \\
  \midrule
\multirow{2}{*}{w4-a16} & w/ clip &115.78&	76.79&	42.04&	10.08&	42.67&	14.63&	13.97&	24.29&	42.53\\
& w/o clip & 31.42&	28.51&	18.75&	9.74&	19.21&	9.80&	9.88&	15.29&	\textbf{17.83} \\
  \midrule
\multirow{2}{*}{w8-a8} & w/ clip & 32.74&	31.41&	17.70&	6.73&	18.02&	7.22&	8.23&	12.01&	16.76\\
& w/o clip & 25.37&	22.04&	15.03&	6.54&	15.74&	6.71&	7.75&	13.13&	\textbf{14.04} \\
  \midrule
\multirow{2}{*}{w8-a16} & w/ clip & 36.31&	22.06&	15.30&	6.18&	14.42&	6.13&	6.16&	11.83&	14.80 \\
& w/o clip & 24.36&	19.66&	14.00&	5.68&	14.56&	5.83&	5.97&	12.76&	\textbf{12.85} \\
  \midrule
 \multicolumn{11}{c}{\textbf{Whisper Base (WER \%)} $\mathbf{\downarrow}$} \\
  \midrule
  & Float &  21.13&	15.09&	12.87&	4.28&	10.36&	4.27&	4.85&	9.75&	10.32\\
  \midrule
\multirow{2}{*}{w4-a8} & w/ clip & 26.86&	21.73&	16.08&	5.26&	13.28&	5.69&	5.71&	11.82&	13.30 \\
& w/o clip & 22.77&	18.95&	13.35&	5.16&	13.13&	5.01&	5.35&	11.23&	\textbf{11.87} \\
  \midrule
\multirow{2}{*}{w4-a16} & w/ clip & 26.77&	20.62&	15.95&	5.67&	14.14&	5.53&	6.11&	11.35&	13.27 \\
& w/o clip & 23.93&	15.86&	13.76&	4.46&	12.64&	4.88&	5.24&	10.48&	\textbf{11.41} \\
  \midrule
\multirow{2}{*}{w8-a8} & w/ clip & 21.69&	15.02&	12.56&	4.22&	10.68&	4.39&	4.89&	9.74&	\textbf{10.40} \\ 
& w/o clip & 28.50&	19.35&	17.48&	9.90&	19.37&	7.42&	9.76&	15.91&	15.96 \\
  \midrule
\multirow{2}{*}{w8-a16} & w/ clip & 20.54&	15.14&	12.54&	4.25&	10.27&	4.27&	4.89&	9.78&	\textbf{10.21}\\
& w/o clip & 21.14&	15.08&	12.74&	4.28&	10.42&	4.23&	4.90&	9.75&	10.32 \\
\bottomrule
 \end{tabular}
\end{adjustbox}
\end{small}
\end{table}

\clearpage
\subsection{Symmetric or Asymmetric Quantization Exeperiments}
\label{appendix:asymmetric}

\begin{table}[h]
\centering
\begin{small}
\captionsetup{font=footnotesize}
\caption{In addition to Table \ref{tab:asymmetric}, per-dataset WER is reported for symmetric vs. asymmetric experiments. Both per-channel and per-group weight quantizations are evaluated. To eliminate other impacts, activations are fixed with $16$-bit per-token quantization.} 
\label{tab:asymmetric_perdataset}
 \scalebox{0.92}{
\begin{adjustbox}{width=\textwidth}
  \begin{tabular}{c c c c c c c c c c c c}
  \toprule
  Granularity & Method & Config. & AMI & Earnings-22 & GigaSpeech & Libri clean & Libri other & SPGISpeech & TED-Lium & Voxpopuli & \textbf{Avg.}\\
  \midrule
 \multicolumn{12}{c}{\textbf{Whisper Base (WER \%)} $\mathbf{\downarrow}$} \\
  \midrule
  & Float &  &  21.13&	15.09&	12.87&	4.28&	10.36&	4.27&	4.85&	9.75&	10.32\\
  \midrule
\multirow{16}{*}{per-channel} & \multirow{2}{*}{SmoothQuant} & sym. & 560.13&	548.78&	554.81&	636.43&	702.38&	702.82&	614.35&	612.87&		616.57\\
& & asym. & 34.14&	55.86&	35.83&	46.82&	47.00&	27.18&	47.05&	39.26&		41.64\\
\cmidrule(r){2-12}
& \multirow{2}{*}{AWQ} & sym. & 55.48&	65.68&	38.30&	42.84&	75.90&	28.94&	37.34&	60.71&	50.65\\
& & asym. & 25.44&	22.36&	14.87&	5.25&	13.74&	6.76&	7.01&	17.52&	14.12\\
\cmidrule(r){2-12}
& \multirow{2}{*}{OmniQuant} & sym. & 31.65&	25.61&	19.18&	6.46&	15.53&	7.78&	6.43&	17.13&		16.22\\
& & asym. & 36.16&	32.18&	20.57&	5.87&	18.09&	8.82&	8.71&	16.62&		18.38\\
\cmidrule(r){2-12}
&\multirow{2}{*}{GPTQ} & sym. & 27.38&	25.61&	16.53&	7.59&	17.56&	7.42&	5.87&	18.80&		15.84\\
& & asym. & 24.18&	21.68&	14.11&	4.74&	13.17&	5.25&	5.64&	12.20&		12.62\\
\cmidrule(r){2-12}
&\multirow{2}{*}{RTN} & sym. & 29.85&	32.33&	17.24&	9.20&	19.78&	8.91&	11.16&	12.81&		17.66\\
& & asym. & 24.33&	19.64&	13.97&	5.40&	13.47&	5.78&	5.69&	12.63&		12.61\\
\cmidrule(r){2-12}
&\multirow{2}{*}{TesseraQ} & sym. & 29.85&	32.33&	17.24&	9.20&	19.78&	8.91&	11.16&	12.81&		17.66\\ 
& & asym. & 23.84&	18.34&	14.09&	5.29&	13.73&	5.65&	5.76&	12.59&		12.41\\
\cmidrule(r){2-12}
&\multirow{2}{*}{QUIK} & sym. & 22.43&	16.99&	13.62&	5.15&	13.24&	5.53&	5.97&	11.63&		11.82\\
& & asym. & 21.46&	16.84&	13.26&	4.58&	11.18&	4.75&	5.28&	11.02&		11.05\\
\cmidrule(r){2-12}
&\multirow{2}{*}{SpQR} & sym. & 22.23&	16.79&	12.94&	4.58&	12.17&	4.81&	5.35&	10.24&		11.14\\
& & asym. & 22.61&	17.14&	12.94&	4.53&	11.71&	4.90&	5.45&	10.19&		11.19\\
\midrule
\multirow{16}{*}{per-group} & \multirow{2}{*}{SmoothQuant} & sym. & 27.35&	17.85&	14.00&	5.04&	12.52&	5.34&	6.54&	10.92&		12.44\\
& & asym. & 23.63&	18.46&	13.56&	4.67&	11.37&	4.95&	5.36&	10.24&		11.53\\
\cmidrule(r){2-12}
& \multirow{2}{*}{AWQ} & sym. & 23.68&	15.87&	14.01&	4.48&	12.24&	4.94&	5.25&	10.56&	11.38 \\
& & asym. & 22.28&	16.79&	12.76&	4.94&	10.98&	4.60&	5.12&	10.15&	10.95\\
\cmidrule(r){2-12}
& \multirow{2}{*}{OmniQuant} & sym. & 23.11&	18.65&	13.79&	4.65&	11.53&	4.92&	5.46&	10.73&		11.61\\
& & asym. & 23.35&	16.45&	14.37&	4.55&	12.82&	4.97&	5.42&	11.41&		11.67\\
\cmidrule(r){2-12}
&\multirow{2}{*}{GPTQ} & sym. & 25.49&	16.68&	13.56&	4.64&	12.03&	4.97&	5.44&	12.25&		11.88\\
& & asym. & 22.34&	15.72&	13.34&	4.43&	10.87&	4.49&	5.12&	10.15&		10.81\\
\cmidrule(r){2-12}
&\multirow{2}{*}{RTN} & sym. & 23.35&	17.53&	13.29&	4.52&	11.66&	4.92&	5.31&	10.78&		11.42\\
& & asym. & 21.70&	16.35&	13.24&	4.99&	12.15&	4.69&	5.10&	10.16&		11.05\\
\cmidrule(r){2-12}
&\multirow{2}{*}{TesseraQ} & sym. & 23.77&	16.64&	13.37&	4.52&	11.63&	4.88&	5.27&	10.83&		11.36 \\ %
& & asym. & 21.90&	16.96&	13.25&	5.01&	12.20&	4.58&	5.15&	10.06&		11.14\\
\cmidrule(r){2-12}
&\multirow{2}{*}{QUIK} & sym. & 21.80&	17.28&	13.76&	4.61&	12.08&	4.76&	5.26&	10.22&		11.22\\
& & asym. & 20.40&	16.03&	12.67&	4.46&	11.36&	4.44&	4.98&	10.86&		10.65\\
\cmidrule(r){2-12}
&\multirow{2}{*}{SpQR} & sym. & 21.62&	16.39&	12.88&	4.42&	11.32&	4.57&	5.24&	10.97&		10.93\\
& & asym. & 22.40&	15.54&	13.21&	4.71&	11.08&	4.49&	5.04&	11.35&		10.98\\
\midrule
\multicolumn{12}{c}{\textbf{Moonshine Base (WER \%)} $\mathbf{\downarrow}$} \\
  \midrule
  & Float &  &  17.07&	17.69&	12.11&	3.26&	8.28&	5.46&	5.24&	10.79&	9.99\\
  \midrule
\multirow{16}{*}{per-channel} & \multirow{2}{*}{SmoothQuant} & sym. & 78.60&	106.55&	94.42&	106.85&	102.91&	95.70&	98.95&	113.42&		99.67\\
& & asym. & 42.40&	59.07&	52.15&	72.20&	70.41&	48.03&	54.19&	65.36&		57.98\\
\cmidrule(r){2-12}
& \multirow{2}{*}{AWQ} & sym. & 19.06&	20.38&	12.96&	3.60&	9.78&	6.08&	5.30&	11.87&	11.13\\
& & asym. & 18.77&	19.28&	12.68&	3.51&	9.29&	5.64&	5.38&	11.49&	10.75\\
\cmidrule(r){2-12}
& \multirow{2}{*}{OmniQuant} & sym. & 19.41&	20.27&	12.73&	3.78&	9.53&	5.81&	5.38&	11.62&		11.07\\
& & asym. & 19.60&	20.62&	12.94&	3.75&	9.53&	5.82&	5.25&	11.83&		11.17\\
\cmidrule(r){2-12}
&\multirow{2}{*}{GPTQ} & sym. & 17.83&	18.82&	12.59&	3.48&	8.89&	5.92&	5.53&	11.26&		10.54\\
& & asym. & 17.84&	18.77&	12.58&	3.52&	8.99&	5.78&	5.52&	11.24&		10.53\\
\cmidrule(r){2-12}
&\multirow{2}{*}{RTN} & sym. & 35.65&	32.53&	17.50&	6.23&	19.64&	11.85&	8.61&	18.15&		18.77\\
& & asym. & 29.19&	29.84&	16.07&	4.78&	14.61&	10.15&	6.83&	17.37&		16.11\\
\cmidrule(r){2-12}
&\multirow{2}{*}{TesseraQ} & sym. & 35.64&	32.65&	17.52&	6.25&	19.56&	11.88&	8.53&	18.18&		18.78\\
& & asym. & 29.02&	29.72&	16.04&	4.75&	14.82&	9.85&	6.92&	17.63&		16.09\\
\cmidrule(r){2-12}
&\multirow{2}{*}{QUIK} & sym. & 19.06&	19.62&	12.63&	3.59&	9.52&	5.79&	5.30&	11.43&		10.87\\
& & asym. & 18.58&	18.59&	12.53&	3.43&	9.15&	6.07&	5.25&	11.46&		10.63\\
\cmidrule(r){2-12}
&\multirow{2}{*}{SpQR} & sym. & 18.11&	18.50&	12.59&	3.39&	8.97&	5.76&	5.23&	11.51&		10.51\\
& & asym. & 17.87&	18.28&	12.46&	3.47&	8.76&	6.08&	5.30&	11.15&		10.42\\
\midrule
\multirow{16}{*}{per-group} & \multirow{2}{*}{SmoothQuant} & sym. &22.44&	22.92&	14.20&	5.51&	15.88&	13.76&	6.34&	12.58&		14.20\\
& & asym. &20.41&	19.18&	12.85&	4.15&	10.51&	7.06&	5.03&	11.97&		11.40 \\
\cmidrule(r){2-12}
& \multirow{2}{*}{AWQ} & sym. & 18.07&	19.39&	12.62&	3.46&	8.89&	5.54&	5.62&	11.57&	10.64\\
& & asym. & 17.94&	18.80&	12.47&	3.37&	8.65&	5.50&	5.50&	10.89&	10.39\\
\cmidrule(r){2-12}
& \multirow{2}{*}{OmniQuant} & sym. & 17.81&	18.41&	12.44&	3.45&	9.04&	5.64&	5.21&	11.11&		10.39\\
& & asym. & 17.88&	18.41&	12.41&	3.49&	8.81&	5.54&	5.60&	11.16&		10.41\\
\cmidrule(r){2-12}
&\multirow{2}{*}{GPTQ} & sym. & 17.60&	18.42&	12.19&	3.45&	8.67&	5.47&	5.45&	11.14&		10.30\\
& & asym. & 17.28&	18.20&	12.24&	3.36&	8.52&	5.59&	5.17&	11.15&		10.19\\
\cmidrule(r){2-12}
&\multirow{2}{*}{RTN} & sym. & 21.78&	23.95&	13.81&	4.52&	11.50&	7.50&	6.60&	13.52&		12.90\\
& & asym. & 19.05&	20.33&	12.71&	3.66&	9.58&	6.07&	5.33&	12.71&		11.18\\
\cmidrule(r){2-12}
&\multirow{2}{*}{TesseraQ} & sym. & 21.88&	24.17&	13.84&	4.58&	11.55&	7.52&	6.69&	13.54&		12.97\\
& & asym. & 18.89&	20.38&	12.66&	3.52&	9.58&	6.17&	5.18&	12.93&		11.16\\
\cmidrule(r){2-12}
&\multirow{2}{*}{QUIK} & sym. & 17.74&	18.80&	12.37&	3.48&	8.96&	5.54&	5.25&	11.05&		10.40\\
& & asym. & 17.88&	18.47&	12.32&	3.45&	8.46&	5.35&	5.26&	11.12&		10.29\\
\cmidrule(r){2-12}
&\multirow{2}{*}{SpQR} & sym. & 17.31&	17.80&	12.37&	3.35&	8.71&	5.47&	5.28&	10.91&		10.15\\
& & asym. & 17.56&	18.28&	12.24&	3.35&	8.55&	5.62&	5.45&	11.07&		10.27\\
\bottomrule
 \end{tabular}
\end{adjustbox}}
\end{small}
\end{table}

\clearpage
\section{Ablation Study} 
\label{appendix:ablation}

To evaluate how calibration data affects quantization performance, we vary both the source and size of the calibration set. Unless otherwise specified, experiments use $4$-bit per-group weight and $16$-bit per-tensor activation quantization.

\textbf{Calibration data sources.} We compare two calibration sources: A held-out subset of $256$ English utterances from Mozilla Common Voice (our default), and a subset of the same size sampled from LibriSpeech Dev. set. Across all eight PTQ algorithms and both models (Whisper Base and Moonshine Base), switching between these calibration sources results in WER variations of less than $1.82\%$, suggesting that most PTQ methods are insensitive to the specific calibration speech corpus, under a good number of sample sizes. 

\begin{table}[h] 
    \caption{Ablation studies on calibration data sources and data sizes. We report the averaged WER across all evaluation datasets for Whisper Base and Moonshine Base models. $4$-bit per-group weight quantization and $16$-bit per-tensor activation quantization is used. (a) For each calibration data source, $256$ samples are used. (b) Using Mozilla Common Voice, calibration sample size varies from $128$ to $512$. RTN does not require any calibration data. Note that weight clipping is not applied to TesseraQ in this set of experiments.}
    \begin{subtable}{.5\linewidth}
      \centering
        \caption{Impact of calibration data source.}
        \scalebox{0.65}{\begin{tabular}{c c c c c}
  \toprule
 \textbf{avg. WER \%} &  \multicolumn{2}{c}{\textbf{Whisper base}} & \multicolumn{2}{c}{\textbf{Moonshine base}} \\
  \cmidrule(r){2-3}
  \cmidrule(r){4-5}
  Calib. Data & Com. Voice & Libri. Dev & Com. Voice & Libri. Dev \\
   \midrule
  SmoothQuant & 12.49 & 12.40 \small{(-0.09)} & 14.20 & 12.38 \small{(-1.82)}\\
  AWQ & 11.41 & 11.80 \small{(+0.39)}& 10.57 & 10.67 \small{(+0.10)}\\
  OmniQuant & 11.59 & 11.72 \small{(+0.13)}& 10.54 & 10.53 \small{(-0.01)}\\
  GPTQ & 11.16 & 11.62 \small{(+0.46)}& 10.28 & 10.40 \small{(+0.12)}\\
  RTN & \--- & \--- & \--- & \---\\
  TesseraQ & 11.42 & 11.37 \small{(-0.05)}& 12.98 & 12.94 \small{(-0.04)}\\ 
  QUIK & 11.30 &11.10 \small{(-0.20)}& 10.40 & 10.33 \small{(-0.07)}\\
  SpQR & 10.94 & 10.85 \small{(-0.09)}& 10.14 & 10.17 \small{(+0.03)}\\
  \bottomrule
  \end{tabular}}
    \end{subtable}%
    \begin{subtable}{.5\linewidth}
      \centering
        \caption{Impact of calibration data size.}
        \scalebox{0.65}{\begin{tabular}{c c c c c c c}
  \toprule
 \textbf{avg. WER \%} &  \multicolumn{3}{c}{\textbf{Whisper base}} & \multicolumn{3}{c}{\textbf{Moonshine base}} \\
  \cmidrule(r){2-4}
  \cmidrule(r){5-7}
  Calib. Samples & 128 & 256 & 512 & 128 & 256 & 512\\
   \midrule
  SmoothQuant & 12.25 & 12.49 & 12.78 & 14.61 & 14.20 & 13.57\\
  AWQ & 11.94 & 11.41 & 97.04 & 10.60 & 10.57 & 18.23\\
  OmniQuant & 11.86 & 11.59 &  12.21 & 10.47 & 10.54 &  10.53\\
  GPTQ & 11.23 & 11.16 & 11.27 & 10.36 & 10.28 & 10.33 \\
  RTN &  \--- & \--- & \--- & \--- & \--- & \---\\
  TesseraQ & 11.37 & 11.42 & 11.37 & 12.94 & 12.98 & 12.94\\
  QUIK & 10.92 & 11.30 & 10.78 & 10.38 & 10.40 & 10.36\\
  SpQR & 10.84 & 10.94 & 10.81 & 10.25 & 10.14 & 10.29\\
  \bottomrule
  \end{tabular}}
    \end{subtable} 
\end{table}


\begin{figure}[!b]
\centering
\begin{subfigure}[b]{0.55\textwidth}
   \includegraphics[width=1\linewidth]{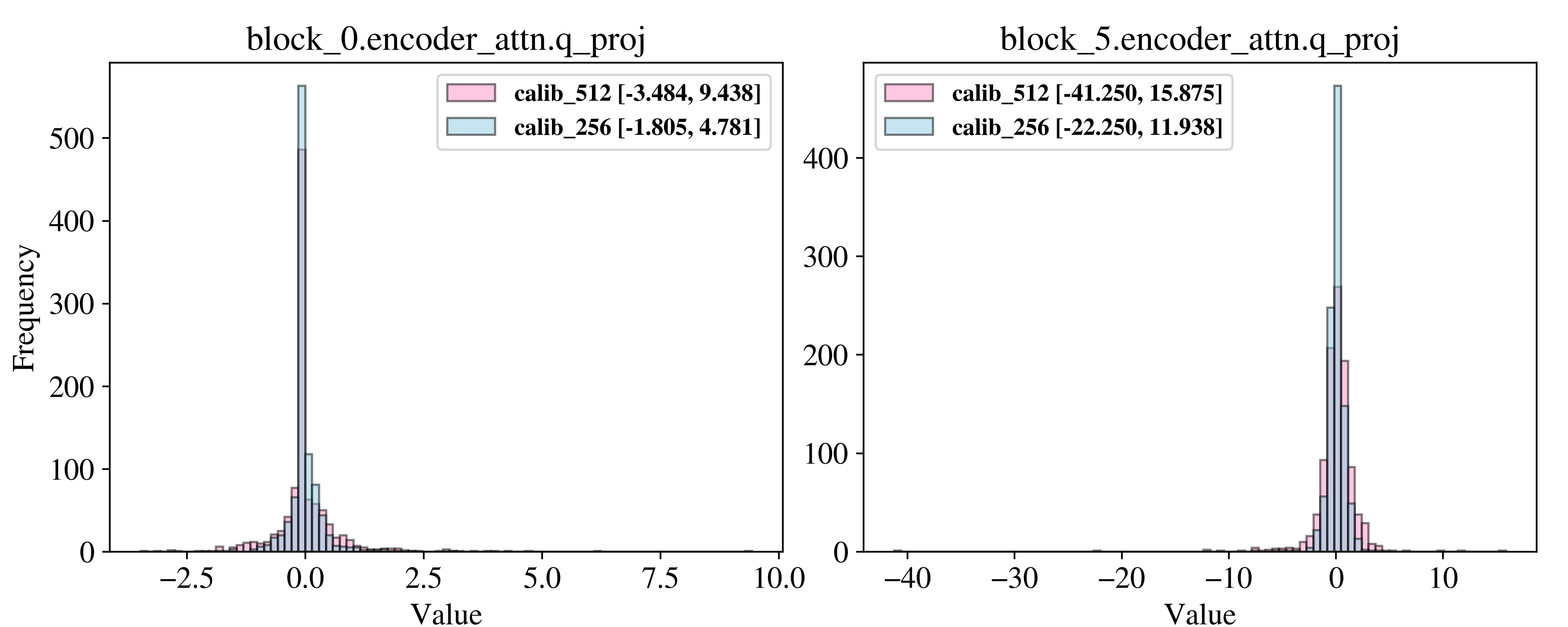}
   \caption{Whisper Base.}
   \label{fig:Ng1} 
\end{subfigure}

\begin{subfigure}[b]{0.55\textwidth}
   \includegraphics[width=1\linewidth]{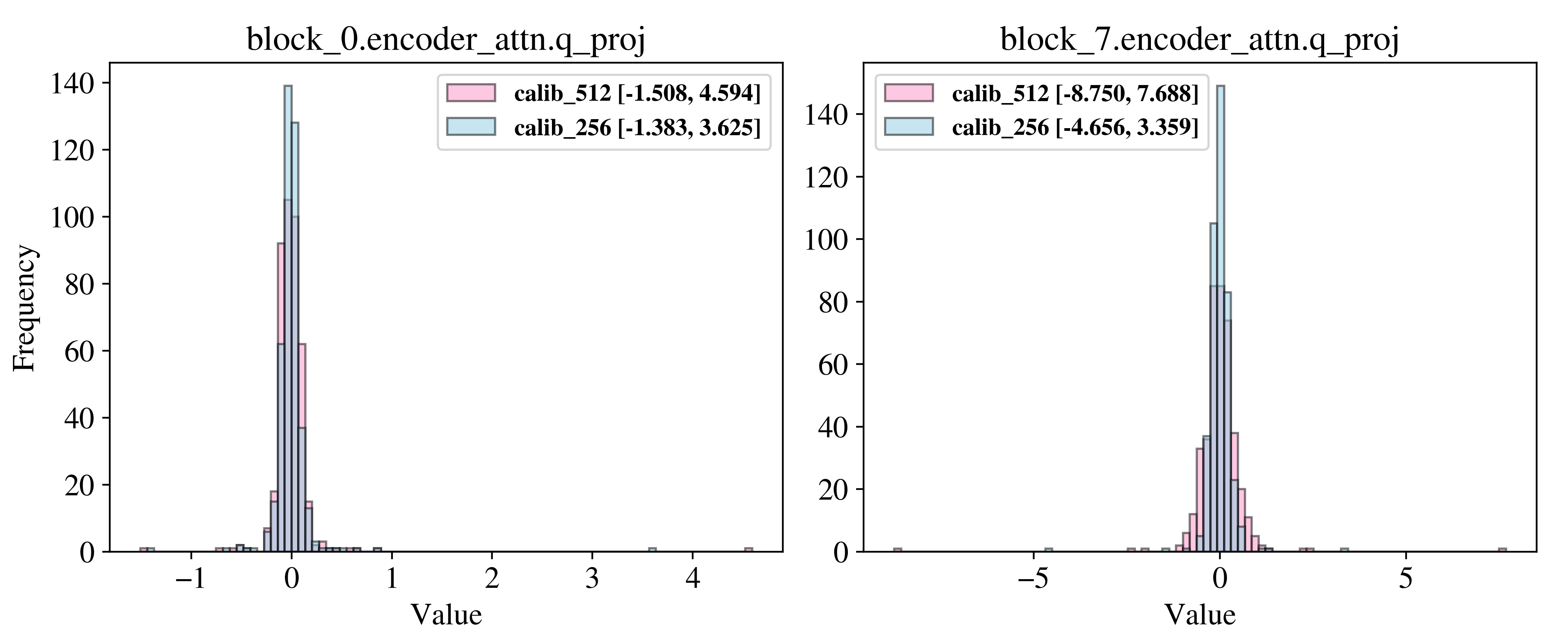}
   \caption{Moonshine Base.}
   \label{fig:Ng2}
\end{subfigure}

\caption{AWQ exhibits a sharp performance drop when sample size increased to $512$. Activation distributions of the first and last block from decoder are examined. The legends denote the calibration sample size, and associated activation data ranges.}
\label{fig:distribution}
\end{figure}

\textbf{Calibration data sizes.} Using Mozilla Common Voice, we further vary the calibration set size across $128$, $256$ and $512$ samples. Increasing the set from $128$ to $256$ samples causes minimal changes in WER across all algorthms. However, when increased to $512$ samples, AWQ exhibits a sharp performance drop. From the block-wise activation distribution shown in Figure \ref{fig:distribution}, a possible reason is that extreme activation outliers dominating the scale computation during $4$-bit weight quantization, which destroys quantization representation for weights. In contrast, other algorithms including reconstruction-based and rounding optimizations, remain stable at $512$ samples. These results suggest that, while most PTQ techniques are generally robust to calibration speech corpus, algorithms sensitive to activation outliers (e.g., AWQ) may require careful tuning of calibration set size to avoid performance degradation. 
\end{appendices}




\end{document}